\documentclass{arxivpreprint}

\usepackage{amsmath,amssymb,bm}
\usepackage[T1]{fontenc}
\usepackage[utf8]{inputenc}
\usepackage[protrusion=true,expansion=false]{microtype} 
\usepackage{float}
\usepackage{enumitem}
\usepackage{ragged2e}
\usepackage{booktabs}
\usepackage{siunitx}
\usepackage{placeins}
\usepackage{animate}
\graphicspath{{figs/}}
\begin{document}
\justifying
\fancyhead[R]{G F Pinton}

\articletype{Paper}

\title{Whole-skull acoustic transparency from a single time-reversal solve for reciprocity-based transducer placement and aperture optimization in transcranial focused ultrasound}

\author{Gianmarco F Pinton$^{1,*}$\orcid{0000-0002-4896-1439}}

\affil{$^1$Lampe Joint Department of Biomedical Engineering, University of North Carolina at Chapel Hill and North Carolina State University, Chapel Hill, NC, USA}

\affil{$^*$Author to whom any correspondence should be addressed.}

\email{gia@email.unc.edu}


\begin{abstract}
\noindent
Transcranial focused ultrasound is limited by the skull, whose thickness, density and
curvature aberrate and attenuate the beam in a position- and target-dependent way, making the
choice of where to couple a transducer for a given deep target a central planning problem. This work shows
that a single full-wave time-reversal solve resolves it. A virtual point source at the target
radiates outward through a computed-tomography (CT) skull model, and by acoustic reciprocity the
time-reversed field gives the transmit coupling of every point on the skull surface at once
(a per-patch skull transparency map) and, in the same pass, the per-element
signals of any array on the model. The recorded field optimizes placement differently for the two
transducer classes. For a phased array, time reversal conjugates the skull aberration,
so the optimum maximizes the delivered energy,
$|p|_{\max}\propto\sqrt{\int_S E\,\mathrm{d}S}$, the surface integral of the delivered-energy density
$E$ over the aperture $S$ (discretely $\sqrt{\sum_i E_i}$). A single element, lacking that
inter-element correction, instead follows a coherent score that mixes delivered energy with phase
coherence across its face. The two classes therefore yield distinct placement optimizers that share a
surface-integral form but differ in integrand and select different windows. Placement, aperture size and orientation are then optimized by
analytically searching this single recorded map in seconds with no further simulation, rather than by a wave
solve per candidate. The same map also bounds the optimal aperture size from the usable, incidence-legal skull. The one outward solve takes of order ten minutes on a single GPU, and the only
remaining solve is a single inward check of the chosen aperture. The map is computed on a
dry-skull micro-CT model for the left dentate nucleus (MNI $[-12,-57,-34]$) with a
$\SI{1}{\mega\hertz}$ heterogeneous Fullwave~2 finite-difference solver~\cite{pinton2021} at \SI{6.16}{} points per
wavelength with bone attenuation. The time-reversed re-emission of a rigid
\ang{120} occipital array refocuses within \SI{0.25}{\milli\meter} of the dentate at
\SI{20.7}{\pascal} per \SI{1}{\pascal} of per-element drive (near-wavelength spot, $-6$~dB full width
at half maximum $1.25\times2.5\times\SI{2.25}{\milli\meter}$). Its delivered-energy-optimal window is
suboccipital, \SI{27}{\milli\meter} from the target at \ang{29.4} incidence, and a 64 mm spherical transducer seated
in this suboccipital basin refocuses on the dentate at $7.9\times$ per-element-drive gain. The aberration correction
 raises the on-target peak $7.5\times$ in pressure ($56\times$ in intensity) over
uncorrected geometric focusing at the same placement, which both mis-steers the focus by
\SI{4.3}{\milli\meter} and defocuses it (confirmed by Fullwave~2 re-simulation). The same field also
measures the aberration's spatial coherence length (a few millimetres), which sets the required
placement and target-localization accuracy. The single-solve prediction reproduces the per-candidate focal
peaks and transverse width to within tolerance. More broadly, the skull transparency map is at once a
visualization and targeting method and a fundamental, frequency-specific characterization of the skull.
From one solve, for any transducer type, it captures the transmitted intensity, the bone transparency,
the aberration coherence length that sets the achievable targeting resolution and dictates whether a
phased array rather than a single element is required, and the regions that are effectively totally
reflecting.
Whereas existing transducer-placement and aberration-correction methods evaluate each candidate
aperture with a separate full-wave solve or rely on single-frequency phase surrogates, the transparency
map is obtained from a single time-reversal solve and returns the delivered-energy-optimal window,
aperture and orientation jointly with the phase aberration and its coherence length, reducing whole-skull
transducer placement to inexpensive post-processing rather than a per-candidate search.
\end{abstract}

\section{Introduction}
Transcranial focused ultrasound (tFUS) is unique among non-invasive techniques in its ability to
deliver acoustic energy to a millimeter-scale focus deep in the brain without a craniotomy \cite{meng2021,blackmore2019,darrow2019}.
Depending on the delivered dose, the same beam can reversibly modulate neural activity
\cite{legon2014,kubanek2018}, transiently open the blood-brain barrier for targeted drug delivery, or thermally
ablate tissue, the last of these already an approved therapy for essential tremor through
magnetic-resonance-guided thalamotomy \cite{elias2016}. Each of these applications depends on
placing a tight, on-target focus through the skull, which is challenging for ultrasound.

Bone is far faster and denser than brain tissue, and its thickness, density, curvature, and trabecular microstructure vary
strongly across the calvarium, so the skull both attenuates the beam and aberrates its phase,
displacing and defocusing the focus by an amount that depends on where the beam crosses the bone. These effects also vary with frequency and transducer geometry. A
large body of work corrects this aberration once a transducer is in place. Per-element phase (and
amplitude) corrections are computed from a subject computed-tomography (CT) scan, either by forward
ray or full-wave propagation \cite{clementhynynen2002,aubry2003,marquet2009} or by time reversal
\cite{fink1992,thomasfink,tanter1998}, and the underlying fields are produced by heterogeneous
full-wave solvers \cite{pinton2009,treeby2010} whose accuracy across implementations has been
established by an international benchmark intercomparison \cite{aubry2022}. A complementary
construction, the radial phase projection \cite{pinton2012}, isolates the skull-induced phase so it
can be transported to an arbitrary transducer surface. All of these answer how the elements should be driven for a given placement, taking the coupling location as
fixed.

A logically prior question is placement. Given a target, where on the skull should the
transducer be coupled, and at what orientation, so that the most energy reaches the target with an
acceptable focus? Because thick, dense or highly curved bone transmits poorly and shifts the focus,
the best window is strongly target-dependent and cannot be read off the scalp geometry alone.
Existing approaches to choosing it fall into three families. The first re-solves the wave equation
for each candidate. One such method \cite{scout} scores a per-candidate acoustic solve and filters by peak negative pressure
and focal volume, while forward planners \cite{babelbrain} model a user-specified
trajectory in detail. These approaches require $O(N_{\rm cand})$ wave solves. The second replaces
the wave physics with a fast geometric surrogate, such as an average reflection-coefficient beamline
\cite{park2019} or scalp- and skull-thickness heuristics
\cite{plantus}. These are inexpensive but discard the aberration physics that sets the true coupling.
The third, closest to this work, uses a single time-reversal solve and acoustic
reciprocity to score placement without per-candidate simulation. A prior reciprocity-based method \cite{park2022}
forms a transmissibility-and-phase score $\psi=|\sum A\,e^{i\Delta\phi}|$ from one reverse solve and
optimizes the pose of a single-element transducer by differential evolution.

Prior reciprocity-based scoring uses one reverse solve to rank a single transducer pose with a
coherent single-element score. That same reverse solve, however, already contains the transmit
coupling of every point on the skull at once. By time-reversal reciprocity the field a virtual point
source at the target radiates to a surface patch equals, up to a known scalar and time reversal
\cite{fink1992}, the field a transducer at that patch would deliver to the target, so one outward
solve yields a dense, whole-skull transparency map rather than the score of a single pose. The
single-element score of \cite{park2022} is, moreover, a coherent sum that combines delivered energy with
phase coherence across the transducer face. For a phased array with per-element phase control the
optimal time-reversal drive conjugates that phase away, and the achievable focal peak collapses to a
closed-form, power-constrained surface integral of the delivered-energy density,
$|p|_{\max}\propto\sqrt{\int_S E\,dS}$, the Cauchy-Schwarz optimum. Placement, aperture size
and shape, and beam orientation then reduce to a moving-window correlation over one precomputed map,
with no iterative search and no per-candidate solve.

That observation is developed here into a validated method. At its core is a reciprocity-based
skull transparency map. A single full-wave time-reversal solve yields the transmit coupling of
the entire external skull surface for a fixed target, rather than of an instrumented window or a single
candidate pose. From it we derive a closed-form, Cauchy-Schwarz-optimal placement objective for a
phased array (the power-constrained surface integral of delivered energy attained by the
time-reversal drive), validated discretely on recorded per-element signals at matched element count
and transmit power. From this one recorded field we read two distinct placement optimizers, the
phased-array delivered-energy integral and the single-element coherent score of \cite{park2022}. They
share the same surface-integral form but differ in integrand, and a full-wave demonstration shows that
their optimal windows differ (energy versus energy-and-coherence).
The same solve also supplies an angular-spectrum focal-spot predictor and a phase-only drive for
arbitrary transducer geometry, via the radial phase projection \cite{pinton2012}. The method is carried out at full-wave, heterogeneous accuracy with one outward solve through
the whole, untruncated skull (no array) on the left dentate nucleus. Multi-element time-reversal
focusing is validated against a rigid \ang{120} occipital array on the same skull-centred model, and
the placement procedure, ground-truthed by inward re-simulation, generalizes to the thalamus and the
dorsal anterior cingulate. The same outward
solve also yields the surface phase-aberration (arrival-time) map and the spatial coherence length of that
aberration, which sets the element pitch and the placement and target-localization tolerance, so one
solve characterizes the skull's transmitted intensity, phase aberration, and coherence for an arbitrary
transducer at the chosen frequency.

\section{Methods}

\subsection{Skull model and target localization}
\label{sec:skullmodel}
The skull is a human dry-skull micro-CT volume (\SI{0.125}{\milli\meter} native pitch)
\cite{halleskull}. Hounsfield units are mapped to a piecewise-linear speed-of-sound
($c=\SI{1540}{\meter\per\second}$ in soft tissue/water rising to \SI{2900}{\meter\per\second} in
cortical bone) and density (\SIrange{1000}{2200}{\kilo\gram\per\meter\cubed}). The volume is block-max downsampled (no interpolation) to an isotropic grid
spacing $\Delta x=\SI{0.25}{\milli\meter}$, $6.16$ points per wavelength (ppw) at \SI{1}{\mega\hertz},
and propagation includes the CT-porosity attenuation model of \S\ref{sec:sim}.

The whole skull is held in a single, skull-centred domain, a $652\times814\times650$ crop
($163\times204\times\SI{162}{\milli\meter}$, $+96$-voxel attenuating boundary) that contains the
entire calvarium, base and facial bones together with the coupling-water margin to seat a
transducer. Inferior and anterior dead space is trimmed. The Fullwave~2 GPU solver stores
fourteen single-precision grids ($\SI{56}{\byte}$/voxel), so the whole skull and a wrapping array
fit in $\approx\SI{28}{\giga\byte}$ on one \SI{48}{\giga\byte} GPU. The same domain serves both the
whole-skull point-source solve (which records the entire bone surface and a decimated volume) and the
occipital-array focusing run (which additionally records the per-element signals), so neither truncation
nor a separate co-registered focusing model is needed.

\begin{figure}[hbt]
\centering
\includegraphics[width=\linewidth]{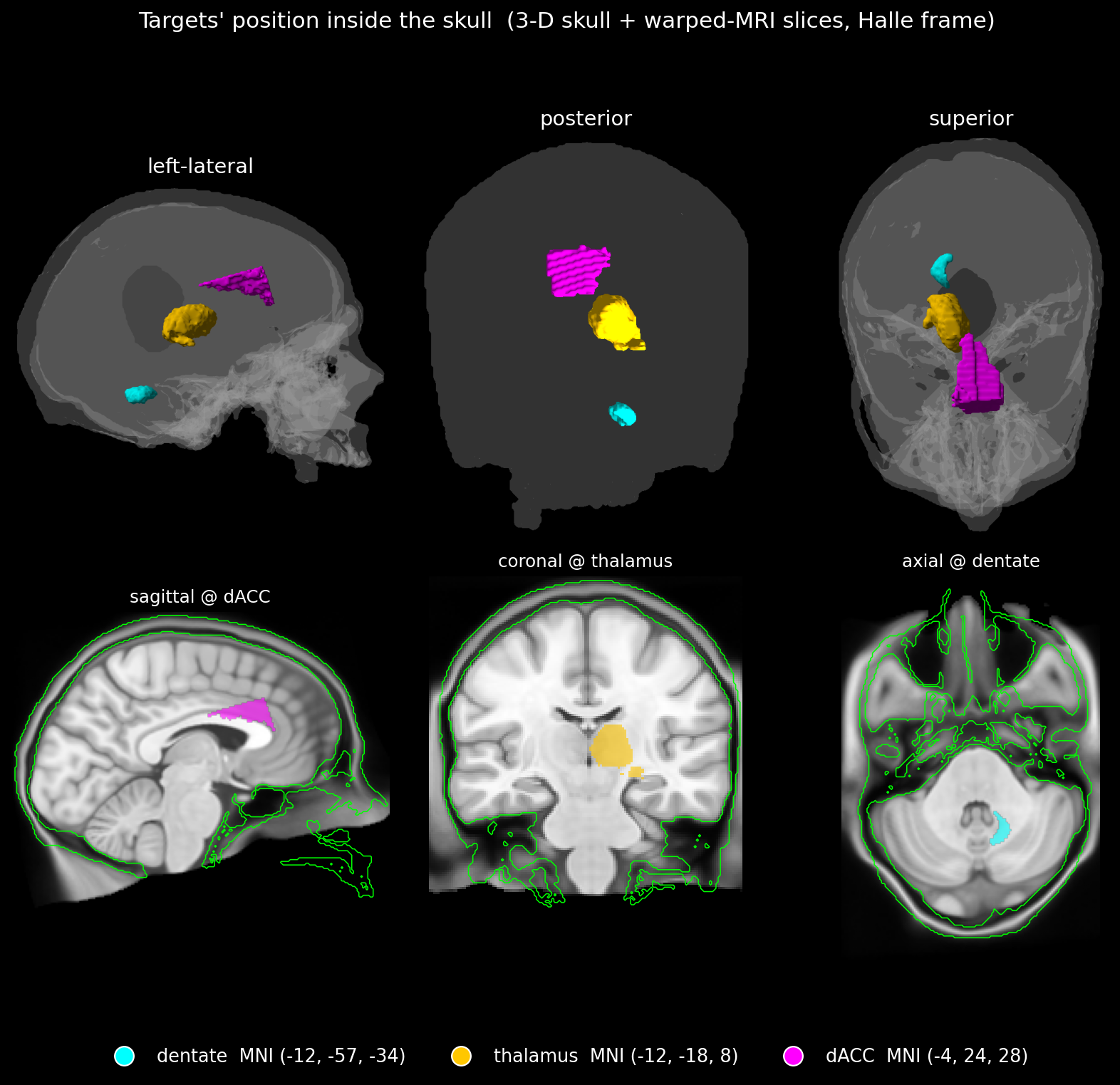}
\caption{The three deep targets in the skull frame (semi-transparent skull, three views), the left
dentate nucleus (cyan, posterior fossa), thalamus (gold, central), and dorsal anterior cingulate
(magenta, anterior). Each is defined in MNI and mapped into the skull by the rigid registration.}
\label{fig:targetsanat}
\end{figure}

The first target is the left dentate nucleus, defined in MNI space at $[-12,-57,-34]$~mm and
mapped into the subject skull via an atlas-to-CT registration (a unified registration
library handles the mappings between the MNI, subject, and grid frames)
\cite{fonov2011,diedrichsen2011}. The map between the simulation and MNI frames is rigid, an orthonormal
rotation anchored at the target. For the whole-skull crop it is obtained in closed form from the
pose and atlas registration and validated against anatomical landmarks (vertex superior, occiput
posterior, left/right correct) before use. The two further deep targets used to demonstrate
generality (Table~\ref{tab:targets}), the thalamus ($[-12,-18,8]$) and dorsal anterior cingulate
($[-4,24,28]$), are localized the same way. Fig.~\ref{fig:targetsanat} shows all three in the skull.

\subsection{Fullwave~2 time-reversal simulation}
\label{sec:sim}
Propagation uses Fullwave~2, a heterogeneous, attenuating full-wave finite-difference time-domain
solver \cite{pinton2009,pinton2021}, on a single GPU (CFL $=0.2$). Absorption follows a CT-porosity model
(\SI{0.4}{\dB\per\mega\hertz\per\centi\meter} in soft tissue/water, rising with porosity in cortical
bone), applied throughout. By default the simulations use a linear \SI{1}{\pascal} drive. A single outward solve sends a
virtual point source at the target through the skull, recording in one pass the field on the
entire external bone surface (the transparency map, \S\ref{sec:surfint}). For focusing, a \ang{120} spherical occipital array (a
single radius of curvature $\approx\SI{71}{\milli\meter}$, 195{,}874 grid sub-elements, seated on
the occipital vault at a few-millimetre coupling standoff, without intersecting bone and with every
element coupling through vault bone while avoiding a foramen path) records the per-element signals in the
outward phase. In the inward phase that recording is time-reversed and re-emitted to refocus at the
target. Because the array has a single radius of curvature, its elements' direct arrivals are coherent, and
the re-emission is windowed to that ballistic cluster with a tapered (Tukey) window. Refocusing quality is reported as
the on-target peak relative to the \SI{1}{\pascal} per-element transmit drive, the focus offset from the
target, and the time-integrated focal $-6$~dB full width at half maximum (FWHM).

\subsection{Surface intensity, transparency, and aberration from the outward solve}
\label{sec:cohmethod}
From the outward phase we form, per grid voxel, the time-integrated intensity
$\mathcal{I}=\sum_t p^2(t)$ (the discrete form of the $\int p^2\,dt$ used to characterize the focal spot) and the peak pressure $p_{\max}=\max_t|p(t)|$ that feeds the surface coupling map below.
On the surface the peak, intensity and arrival time are evaluated only within the physical
direct-arrival (ballistic) window $t\le 1.12\,|\mathbf{x}-\mathbf{x}_t|/c_0$, so the coupling
map is the ballistic transmission and is clean of the later reverberation coda. The external calvarial
surface is extracted as the set of bone/tissue boundary voxels whose outward neighbour is
tissue and whose inward neighbour is bone. The field is sampled at a small standoff just
outside each surface patch. For each patch $i$ we record its position, the distance $r_i$ to
the target, the peak pressure, the peak intensity $I_i=p_{\max,i}^2/(2\rho c)$ (with $p_{\max,i}$ the peak pressure at patch $i$, and $\rho$, $c$ the density and sound speed of the standoff medium just outside the
bone), and the true local
surface normal (from the smoothed bone-occupancy gradient). The raw surface intensity $I_i$,
dominated by proximity to the target, is the coupling quantity used for placement (\S\ref{sec:costfn}).

The same outward record carries the phase. At each surface patch we take the direct-arrival time
$t_{\mathrm{arr}}(\mathbf{x})$ as the parabolic-refined peak of the analytic-signal (Hilbert) envelope
within the same ballistic window, and form the aberration delay
$\Delta t(\mathbf{x})=t_{\mathrm{arr}}(\mathbf{x})-|\mathbf{x}-\mathbf{x}_t|/c_0$, zero-median centred
over the coherent surface (the per-element transmit-delay correction, mapped in \S\ref{sec:delaymap}).
Two quantities characterize the aberration. Its magnitude is the local RMS delay
$\sqrt{\langle\Delta t^2\rangle}$ over a surface neighbourhood, mapped across the skull. Its
spatial coherence is estimated in two independent ways as a function of patch separation $d$,
(i) the autocorrelation of the scalar delay map,
$C_1(d)=\langle\Delta t(\mathbf{x})\,\Delta t(\mathbf{x}{+}\mathbf{d})\rangle/\langle\Delta t^2\rangle$,
and (ii) the zero-lag normalized cross-correlation of the recorded waveforms themselves,
$\rho_0(d)=\langle\hat{w}_i\!\cdot\!\hat{w}_j\rangle$, where each unit-energy direct-arrival pulse
$\hat{w}$ is first sub-sample aligned (by an FFT phase shift) to its spherical arrival so the residual
decorrelation is the aberration alone (a van~Cittert--Zernike coherence~\cite{mallart1991}). Both equal unity at $d=0$ by
construction, and the coherence length is defined as the smallest separation at which the
coherence falls to $e^{-1}$,
\begin{equation}
L \;=\; \min\{\, d \ge 0 \;:\; C_1(d) \le e^{-1} \,\},
\label{eq:coherence_length}
\end{equation}
with the equivalent estimate from $\rho_0(d)$ used as an independent cross-check. This phase analysis, like the
amplitude (transparency) map, is obtained from the single outward solve. Results are reported in
\S\ref{sec:coherence}.

\subsection{Time-reversal reciprocity and transducer placement}
By acoustic reciprocity for the linear, lossy, heterogeneous wave equation, the Green's
function is symmetric,
\begin{equation}
G(\mathbf{x}_{\text{target}},\mathbf{x}_{\text{surf}};\omega)=
G(\mathbf{x}_{\text{surf}},\mathbf{x}_{\text{target}};\omega).
\label{eq:reciprocity}
\end{equation}
Hence, at each frequency, the outward field
measured at a surface patch equals (up to phase conjugation, i.e.\ time reversal) what a transducer at
that patch would deliver to the
target. For placement (maximizing delivered energy) the relevant quantity is
the raw surface intensity $I_i$ (\S\ref{sec:costfn}).

To place a single focused-bowl transducer we (i) weight each patch by incidence,
$w_i=\cos^2\theta_i$ with $\cos\theta_i=\langle \hat{\mathbf{r}}_i, \hat{\mathbf{n}}_i\rangle$
(the radial unit vector $\hat{\mathbf r}_i$ pointing from the target outward to patch $i$, versus the
true outward surface normal $\hat{\mathbf n}_i$, the same $\hat{\mathbf r}_i$ reappearing in the radial
projection and point-spread of \S\ref{sec:surfint}),
and hard-reject patches beyond a maximum incidence (\ang{30}, near the water-to-bone longitudinal critical angle, beyond which the longitudinal transmission the fluid solver models gives way to shear-mode conversion \cite{clement2004,white2006}), (ii) aggregate the incidence-weighted
coupling over the transducer footprint (patches within the aperture radius), and (iii) choose the
window that maximizes this aggregate. The transducer axis is set from the chosen window to the
target, and the apex is placed one focal length from the target so the focus lands on the
target. The single-solve transparency map makes step (ii) free to evaluate for arbitrarily
many candidate placements. Only final verification of inter-element coherence requires a
re-simulation.

For a multi-element aperture the same map populates the chosen window directly, its transducer
footprint filled with a densely-sampled (by default a $\le\lambda/2$ pitch), surface-conformal element set,
each element oriented toward the target, and the per-site incidence-weighted coupling
$w_i I_i=I_i\cos^2\theta_i$ supplies an apodization weight. A dense, fully-populated aperture maximizes delivered energy and suppresses grating lobes. Where the channel count is hardware-limited, the same
per-site ranking instead can select or apodize a reduced element set, which will reduce the fill factor and
raise the side-lobe floor for fewer channels \cite{pernot2003}.

\subsection{Fair validation of placement and focusing}
\label{sec:validation}
The same outward solve also records the full pressure signal
$r_i(t)$ reaching each candidate element site of the occipital array from the target point source. By
reciprocity $r_i(t)$ (a recorded pressure trace, not the patch-to-target distance $r_i$) is the
element-to-target channel response, so the on-target field produced by driving a
set of elements with drives $d_i$ is the on-target pressure
\begin{equation}
s(t)=\sum_i (r_i * d_i)(t),
\label{eq:superpose}
\end{equation}
where $*$ is temporal convolution and each driven element re-radiates through its own channel
response $r_i$. The placement and
focusing comparisons can therefore be evaluated directly on measured data, with no further
solve, at matched element count $N$ and matched transmit power
$\sum_i\lVert d_i\rVert^2=1$ (with $\lVert d_i\rVert^2=\int d_i^2\,dt$). Two quantities follow. The first is placement where under full time
reversal ($d_i\propto r_i(-t)$) every element co-phases on target, so the power-normalized
on-target peak obeys
\begin{equation}
|s|_{\max}\;\propto\;\sqrt{\sum_i E_i},\qquad E_i=\int r_i^2\,dt,
\label{eq:trpeak}
\end{equation}
with $E_i$ the per-element delivered energy, independent of aperture geometry. Placement quality is thus exactly the
total delivered energy $\sum_i E_i$, which the transparency map ranks per site. This work compares the
transparency-selected aperture against transparency-blind baselines (uniform tiling of the
same window) at equal $N$. The second is
focusing. At a fixed placement we delay-and-sum the measured traces with two
delay laws, the medium's true (aberration-corrected) delays $\tau_i$ versus the
homogeneous-medium geometric delays $\lvert\mathbf{x}_i-\mathbf{x}_t\rvert/c_0$, with $\mathbf{x}_i$ the
$i$th element center, $\mathbf{x}_t$ the target point, and $c_0$ the constant homogeneous reference sound
speed (the soft-tissue/water value of \S\ref{sec:skullmodel}). This isolates
the skull-aberration loss that time reversal recovers. Focal volume and off-target
lobes are then measured by a full-wave inward re-simulation of the chosen aperture under each
drive law. The data-driven metrics rapidly quantify the on-target peak and the
placement/focusing gains, and the re-simulation confirms them more rigorously.

\subsection{Single-solve surface-integral placement}
\label{sec:surfint}
The validation above is a special case of a general result that makes placement
optimization essentially free. Working in the frequency domain (the
$e^{-i\omega t}$ transform convention, under which the recorded phasor of a causal arrival carries
$\arg G\approx-kr$, with $k=\omega/c_0$ and $r$ the radial propagation distance from the target to the
recording point), a transducer
occupying a surface $S$ with complex drive $u(\mathbf{x})$ produces at the target point $T$
($T\equiv\mathbf{x}_{\text{target}}$, written $\mathbf{x}_t$ in \S\ref{sec:validation}) the field
\begin{equation}
p(T)=\int_S u(\mathbf{x})\,G(T,\mathbf{x})\,dS,
\label{eq:forward}
\end{equation}
where $G(T,\mathbf{x})$ is the complex field the single outward solve records at $\mathbf{x}$
(by reciprocity, $G(T,\mathbf{x})=G(\mathbf{x},T)$). At each frequency, for a fixed transmit power
$\int_S|u(\cdot,\omega)|^2\,dS=P(\omega)$, Cauchy--Schwarz gives
\begin{equation}
|p(T,\omega)|_{\max}=\sqrt{P(\omega)\int_S|G(\omega)|^2\,dS},
\label{eq:cauchy}
\end{equation}
attained by the phase-conjugate (time-reversal) drive $u\propto G^*$. This same drive co-phases
every frequency at $T$, making $p(T,\omega)$ real and positive ($\arg p(T,\omega)=0$ for all
$\omega$). The frequency components therefore add coherently at the focusing instant, and the
time-domain peak is the in-phase sum $\int|p(T,\omega)|_{\max}\,d\omega$. Integrating the
per-frequency power constraint over the pulse band (Parseval) makes the placement objective the
surface integral of the broadband delivered-energy density $E(\mathbf{x})=\int|G(\mathbf{x},\omega)|^2 d\omega$,
\begin{equation}
|p(T)|_{\max}\;\propto\;\sqrt{J(S)},\qquad J(S)=\int_S E(\mathbf{x})\,dS,
\label{eq:surfint}
\end{equation}
a single precomputed map. The discrete $\sqrt{\sum_i E_i}$ of \S\ref{sec:validation} is exactly
the sampled $\sqrt{J}$ (by Parseval, $E_i=\int r_i^2\,dt=\int|G_i(\omega)|^2\,d\omega=E(\mathbf{x}_i)$).
The single-frequency density $|G(f_0)|^2$ at the center frequency $f_0=\SI{1}{\mega\hertz}$ is only a proxy for $E$ (here
they correlate at $r\!=\!0.89$ once ballistic-windowed). This work uses the direct-arrival (ballistic)
energy density (the peak intensity $I=p_{\max}^2/(2\rho c)$ on the dense surface, or the
ballistic-windowed energy on the array) because the full-record time-integral is
reverberation-contaminated, which can produce coda-dominated errors in the window-selection criteria. Sliding the aperture over the candidate
sphere is one moving-window (spherical) correlation, so every candidate position, aperture
size/shape, and beam orientation is evaluated by summation, with no further wave solve. When a
target's best window abuts a no-bone opening (e.g. the foramen magnum, for the dentate), the same
summation is restricted to the transducer's kept aperture, the elements whose ray to
the target crosses bone, dropping the rest element-by-element (\S\ref{sec:transparency}), and is
maximized over the full transducer pose (window centre, aim, and standoff). Optimizing over the pose, with
the per-element exclusion, refines the seat for every target. For the dentate it keeps the foramen-clean
suboccipital window that a whole-transducer foramen-free constraint would have abandoned for a distal
occipital one. For the foramen-free thalamus and dACC it trades a little window-centre coupling
for better whole-aperture incidence and fewer dropped rays. 

Candidate transducers at a different focal radius $R$ are handled by the
radial projection \cite{pinton2012}. Beyond the bone the aberrated wave propagates radially, so
the recorded phasor $G$ (carrying the retarded radial phase $\arg G\approx-kr$ noted above) is carried
between concentric spheres about $T$ by the per-ray operator
\begin{equation}
G(R\hat{\mathbf r})=G(r_1\hat{\mathbf r})\,\frac{r_1}{R}\,e^{-ik(R-r_1)},
\label{eq:radproj}
\end{equation}
where $r_1$ is the radius of the sphere on which $G$ is recorded and $R$ the candidate focal radius
along ray $\hat{\mathbf r}$ (amplitude spreading plus radial phase), an $O(M)$ pass that conserves the
product (radius)$^2|p|^2$ between concentric spheres along each ray.
From the same field one also obtains two further quantities. The first is the practical
phase-only (equal-amplitude) focal peak, with $|S|$ the aperture area,
\begin{equation}
|p(T)|_{\text{eq}}=\sqrt{\frac{P}{|S|}}\int_S|G|\,dS,
\label{eq:phaseonly}
\end{equation}
which sits a factor $1/\sqrt{1+\mathrm{CV}^2(|G|)}$ below the optimum, where
$\mathrm{CV}(|G|)=\mathrm{std}_S(|G|)/\mathrm{mean}_S(|G|)$ is the coefficient of variation of the
amplitude over $S$. The second is an order-of-magnitude focal-spot predictor, at a field point $T'$
near the focus, from the angular-spectrum point-spread
\begin{equation}
p(T')\approx\sum_i|G_i|^2\,e^{ik\,\hat{\mathbf r}_i\cdot(T-T')},
\label{eq:psf}
\end{equation}
the sampled form of $\int_S$ over surface patches $i$, with $G_i=G(T,\mathbf{x}_i)$ and
$\hat{\mathbf r}_i$ the ray direction at patch $i$.

This reformulates the cost of placement optimization. Where a per-candidate-simulation method
\cite{scout} pays $O(N_{\rm cand})$ full-wave solves, Eq.~\eqref{eq:surfint} needs one outward solve per target, the
same solve already required for time-reversal focusing and of order ten minutes on a single GPU.
Placement, aperture and orientation are then post-processing on the precomputed map in seconds,
independent of the number of candidate placements.
The only residual wave solve is the final inward verification of the single chosen
aperture (\S\ref{sec:validation}), not a per-candidate loop. Two caveats, quantified in Results (\S\ref{sec:surfintresult}), bound the focusing claims. The single-frequency phase-conjugate-versus-geometric ratio overstates the broadband aberration gain, and the single-frequency PSF overestimates the axial depth of field. Nevertheless, placement, which depends only on the
energy integral, is insensitive to both.

\subsection{Array versus single-element apertures and the energy-coherence trade-off}
\label{sec:arrayvssingle}
Equation~\eqref{eq:surfint} assumes the drive $u(\mathbf{x})$ is free across the aperture, as in a
phased array with per-element amplitude and phase control. A single-element
transducer instead radiates its whole face with one waveform and a fixed geometric-focusing
curvature $\phi_\mathrm{geo}(\mathbf{x})=+k\,r(\mathbf{x})$, $r=|\mathbf{x}-T|$ (the conjugate of the
homogeneous-medium retarded phase $\arg G\approx-kr$, so a homogeneous medium would focus at $T$). It
cannot correct the skull aberration across its own face. One recorded field $G$ therefore
defines two distinct placement optimizers, the array optimizer (maximize delivered energy) and
the single-element optimizer (maximize the coherent energy-and-phase score). They share the
surface-integral form but optimise different integrands,
\begin{equation}
J_\mathrm{arr}(S)=\int_S |G|^2\,dS \quad(\text{array, phase corrected}),\qquad
J_\mathrm{sgl}(S)=\Big|\int_S G\,e^{i\phi_\mathrm{geo}}\,dS\Big|^2=\Big|\int_S |G|\,e^{i\Delta\phi}\,dS\Big|^2,
\label{eq:arrvssgl}
\end{equation}
with $\Delta\phi(\mathbf{x})=\arg G+\phi_\mathrm{geo}=\arg G+kr$ the residual aberration phase, the
departure of the recorded phase from the homogeneous-medium value $-kr$, hence $\Delta\phi\to0$ in a
homogeneous medium. The array
optimum maximises delivered energy (the phase is conjugated away, $u\propto G^*$). The
single-element optimum maximises a coherent sum that rewards energy and phase
coherence across the face. Their ratio is the spatial coherence factor
\begin{equation}
\gamma(S)=\frac{\big|\int_S|G|\,e^{i\Delta\phi}\,dS\big|}{\int_S|G|\,dS}\in[0,1],
\label{eq:gamma}
\end{equation}
with $\gamma=1$ for a phase-flat window. An array's per-element delays restore $\gamma\to1$ and
recover the full $\int_S|G|\,dS$. A single element, lacking that correction, delivers only the
coherent amplitude $\gamma\int_S|G|\,dS$, a fractional penalty $1-\gamma$. This penalty is the
aberration loss corrected by the array (\S\ref{sec:crossover}). Since $\gamma$ varies over the skull, the two
optimizers are maximised at different windows. The array optimizer places for maximum
transparency (delivered energy), whereas the single-element optimizer should trade some transparency for a
flatter-phase window. That two optimizers read from one solve select different optimal windows is itself
a result, not a corollary of a single blended objective. This crossover is consistent with the radial
phase projection \cite{pinton2012}, which isolates exactly the residual aberration $\Delta\phi$ that the
phased array cancels and the single element instead incurs as loss.

The single-element limit $J_\mathrm{sgl}$ matches the placement score of prior reciprocity-based work
\cite{park2019,park2022} ($\psi=|\sum A\,e^{i\Delta\phi}|$, transmissibility $A=|G|$), likewise
evaluated from one time-reversal solve. Our contribution is the complementary phased-array optimizer
$J_\mathrm{arr}$, where the objective is pure delivered energy, and the observation that one recorded
field carries both optimizers in a common surface-integral form, with the achievable on-target
peak following the same surface integral in either,
\begin{equation}
|p(T)|_{\max}\propto\sqrt{P\,J_\mathrm{arr}(S)}\ \ (\text{array}),\qquad
|p(T)|_{\max}\propto\sqrt{\frac{P}{|S|}\,J_\mathrm{sgl}(S)}\ \ (\text{single element}),
\label{eq:unify}
\end{equation}
the $1/|S|$ being the equal-amplitude power normalization of \S\ref{sec:surfint}, with $J$ interpolating
from the coherent single-element form $J_\mathrm{sgl}$ to the energy phased-array form $J_\mathrm{arr}$
as independently-phased channels are added. Of the placement families of \S1, our objective keeps the full-wave physics of the per-candidate planners \cite{scout,babelbrain} at the single-solve cost of the geometric surrogates \cite{park2019,plantus}. This predicted crossover, and the dominant aberration penalty that accompanies it, is tested against a
full-wave ground truth in the Results (\S\ref{sec:crossover}).

\subsection{The placement cost function}
\label{sec:costfn}
Collecting the preceding results, the placement objective to maximise is the array optimizer, an
incidence-weighted, legality-masked refinement $J_\mathrm{w}$ of the bare delivered-energy integral $J$
of Eq.~\eqref{eq:surfint},
\begin{equation}
J_\mathrm{w}(S)=\int_S \cos^2\theta(\mathbf{x})\,E(\mathbf{x})\,\Lambda(\mathbf{x})\,dS,\qquad
p_{\max}\propto\sqrt{P\,J_\mathrm{w}(S)} ,
\label{eq:costfn}
\end{equation}
which gives the achievable on-target peak pressure at fixed transmit power $P$ and collapses to $J$ at
normal incidence. The legality mask $\Lambda$ is a binary $\{0,1\}$ no-go mask that zeroes off-limit
windows before scoring, the steeply incident paths beyond \ang{30}, the paranasal sinuses and other
intracranial air cavities, the foramina, and the orbits. The integrand $E=|G|^2$ is the raw,
direct-arrival, broadband delivered-energy density of \S\ref{sec:surfint} (the energy form
$J_\mathrm{arr}$ of Eq.~\eqref{eq:arrvssgl}, with a single element using the coherent integrand instead,
\S\ref{sec:arrayvssingle}). Incidence enters twice. A soft $\cos^2\theta$ weight on the true surface
normal grades obliquity continuously, and the hard \ang{30} cutoff in $\Lambda$ excises the grazing
tail and fixes the access bound of Table~\ref{tab:targets}. Evaluating $J_\mathrm{w}$ over a sliding
aperture footprint is one moving-window correlation (\S\ref{sec:surfint}).

The integrand follows the drive model of \S\ref{sec:arrayvssingle}, the energy form $J_\mathrm{arr}$ for
a phased array and the coherent sum $J_\mathrm{sgl}$ for a single element, with the window shift between
them growing with the target's aberration (\S\ref{sec:crossover}). Amplitude apodisation (full time
reversal versus phase-only) is a minor $5$--$13\%$ refinement that does not change the window ranking.

Equation~\eqref{eq:costfn} returns the on-target peak only. A deployable cost extends it
hierarchically rather than as a single blended scalar, because the additional terms are of
different kinds. Hard regulatory and anatomical gates ($\Lambda$, including a per-aperture
skull-surface dose/heating limit) are applied first. A registration de-rating
$D(S)=\mathbb{E}_\Delta[J_\mathrm{w}(S;T{+}\Delta)]/J_\mathrm{w}(S;T)$ over the localisation-uncertainty ball
($\sigma\!\approx\!1$--$3$~mm, re-scored on the same map, exact for $\Delta$ well below the focal
size) rewards forgiving windows and flags knife-edge ones, such as the narrow-access dACC. The competing soft objectives (focal volume,
off-target/grating lobes, peak skull-surface intensity) are then exposed as a Pareto front among
near-optimal windows, scored from the angular-spectrum point-spread and, for the single chosen
aperture, one inward re-simulation. The energy core, incidence weight and (approximate) $D$ are all
computed from the single map, with only the final selectivity/safety check needing a wave solve.

\subsection{Optimal aperture size}
\label{sec:aperturesize}
The map also bounds the optimal aperture size, but the logic is the opposite of a peak-finding contour, since
focusing admits no interior size optimum. The focal peak
$p_{\max}\propto\sqrt{\int_S \cos^2\theta\,E\,\Lambda\,dS}$ grows monotonically with aperture and the focal width
($\sim\!\lambda f_\#$) shrinks with it, so a larger aperture is always better, until it runs off the
usable skull. The optimum is therefore set by a bound, not an extremum. The aperture grows until it leaves
the incidence-legal, well-coupled window (the access region of Table~\ref{tab:targets}) or reaches the practical
fast-focusing floor (grating lobes, fabrication, $f_\#\!\gtrsim\!0.5$). Whether the binding constraint is the legal window or the focusing floor depends on how each target's available access compares to a focusing transducer's fixed solid-angle demand. The per-target regimes and numbers are in Results (\S\ref{sec:placement}, Table~\ref{tab:aperture}). We deliberately
do not size the aperture from the proximity-weighted energy contour, which measures only how fast the
delivered energy falls off around the nearest patch, a proximity scale almost independent of the
target, rather than the extent over which coherent focusing remains possible.

\section{Results}

The results proceed in four stages. Full-wave transcranial refocusing is first demonstrated on the
dentate with a realistic occipital array (\S\ref{sec:refocus}). The surface coupling,
phase-aberration, and coherence maps that the single outward solve yields are then read
(\S\ref{sec:transparency}--\S\ref{sec:coherence}). Those maps are next used to select the window and
aperture for all three deep targets and verify each by inward full-wave re-simulation
(\S\ref{sec:placement}--\S\ref{sec:targetfoci}). Finally, this work tests the two predictions that distinguish
the method, the array-versus-single-element crossover against a full-wave ground truth
(\S\ref{sec:crossover}) and the claim that the single-solve surface integral reproduces per-candidate
simulation at a fraction of the cost (\S\ref{sec:surfintresult}).

\subsection{Realistic transcranial refocusing on the dentate}
\label{sec:refocus}
\begin{figure}[t]
\centering
\includegraphics[width=\linewidth]{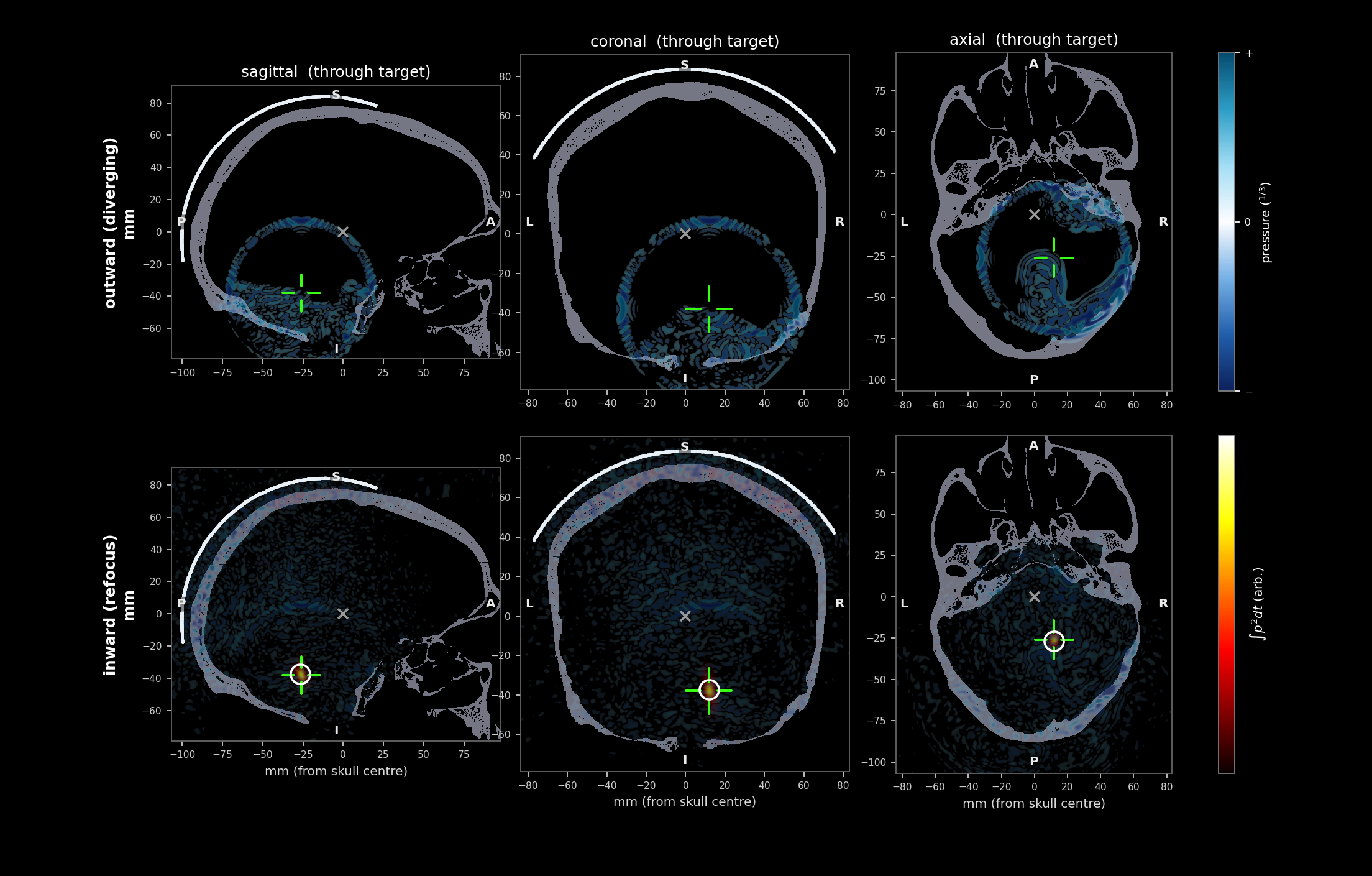}
\caption{Occipital-array time reversal at \SI{1}{\mega\hertz} on the single skull-centred,
attenuating domain (anatomical slices through the target). In the outward phase (top) a point source
at the left dentate nucleus radiates through the skull as a cool (white/blue) diverging wavefront (pale
arc = the \ang{120} rigid occipital array on the posterior-superior vault). In the inward phase (bottom)
the time-reversed re-emission converges to a focus on the dentate. The hot background is the cumulative
intensity $\int p^2dt$ (focal energy) with the cool converging wavefront overlaid. Pressure is
$^{1/3}$-power-compressed for display. Green crosshair = target, white ring = achieved focus.}
\label{fig:propagation}
\end{figure}

\begin{figure}[t]
\centering
\includegraphics[width=\linewidth]{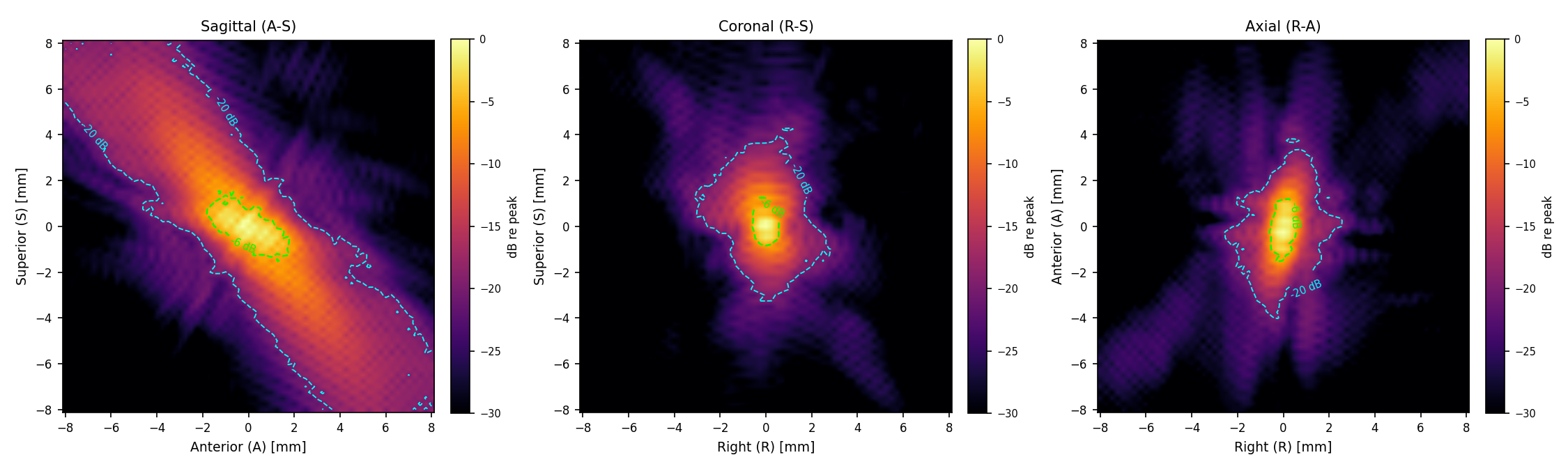}
\caption{The \SI{6.16}{}-ppw transcranial focal spot at the dentate, from the attenuated inward
re-emission of the rigid occipital array at \SI{1}{\pascal} per-element transmit drive. Three
orthogonal slices through the focal peak (sagittal, coronal, and axial planes, inferno colormap)
within the focal box, displayed in dB relative to the peak with axes labeled in mm and centered
on the focus. The \SI{-6}{\decibel} (FWHM) contour is drawn in white and the \SI{-20}{\decibel}
focal-spot skirt in cyan. Focal peak \SI{20.7}{\pascal} ($20.7\times$ gain). FWHM and offset in
text (\S\ref{sec:refocus}).}
\label{fig:focus}
\end{figure}

Figure~\ref{fig:propagation} shows the two phases of the occipital-array focusing run, the diverging
wave from the dentate filling the skull and the time-reversed convergence back onto the target
(animated in Fig.~\ref{fig:propmovie}). With
the rigid occipital array driven at \SI{1}{\pascal} per-element transmit drive and the skull's attenuation
included, the re-emission refocuses on the dentate within \SI{0.25}{\milli\meter} of the target and
reaches a focal peak of \SI{20.7}{\pascal}, a $20.7\times$ transcranial focal-pressure gain through
solid occipital bone.
The focus is near-wavelength ($-6$~dB FWHM $1.25\times2.5\times\SI{2.25}{\milli\meter}$). The focal
spot is shown in Fig.~\ref{fig:focus}. The drive is a short broadband pulse, a roughly two-cycle
super-Gaussian-windowed sinusoid centred at \SI{1}{\mega\hertz}, rather than the long near-continuous-wave
tone burst typical of neuromodulation. The broadband pulse sharpens the direct (ballistic) arrival and
separates it from the later reverberation coda, which is what makes the arrival-time and coherence maps
well defined. The \SI{1}{\mega\hertz} centre frequency is chosen for finer focal and arrival-time
accuracy than the lower neuromodulation frequencies.

\begin{figure}[t]
\centering
\animategraphics[controls,loop,width=\linewidth]{6}{figs/movie_frames/frame-}{0}{59}
\caption{Animated full-wave transcranial propagation through the skull, using the same
skull-centred anatomical slices as Fig.~\ref{fig:propagation} (grey = skull, cyan = \ang{120} rigid
occipital array, green $+$ = dentate target). The movie plays the outward diverging wavefront
sweeping out from the dentate to the full transducer, followed by the inward time-reversed converging
re-emission refocusing onto the target. Each phase uses a fixed pressure colour scale (RdBu). Playable in Acrobat and compatible PDF viewers. Otherwise a single representative frame is shown.}
\label{fig:propmovie}
\end{figure}

\subsection{Surface coupling and bone transmission}
\label{sec:transparency}
The single outward solve yields the acoustic coupling at every surface point. For the dentate (\num{3997167} surface patches spanning the complete calvarium,
base and facial bones) the field is strongly peaked on the occipital and suboccipital bone nearest the
target, the shortest path and least geometric spreading, with a peak surface pressure of
\SI{85}{\milli\pascal} (this maximum is the basal foramen-magnum leak discussed below, not a placeable
window). The emergence is target-specific, concentrated on the occiput for the eccentric dentate,
and distributed more broadly over the calvarium for the centrally-located thalamus and the anterior
dACC. Because the source amplitude is arbitrary (matched across targets), the spatial
pattern and cross-target ranking, not the absolute level, are the meaningful output (raw maxima 85, 49, 44 mPa for dentate, thalamus, dACC).

\begin{figure}[t]
\centering
\includegraphics[width=\linewidth]{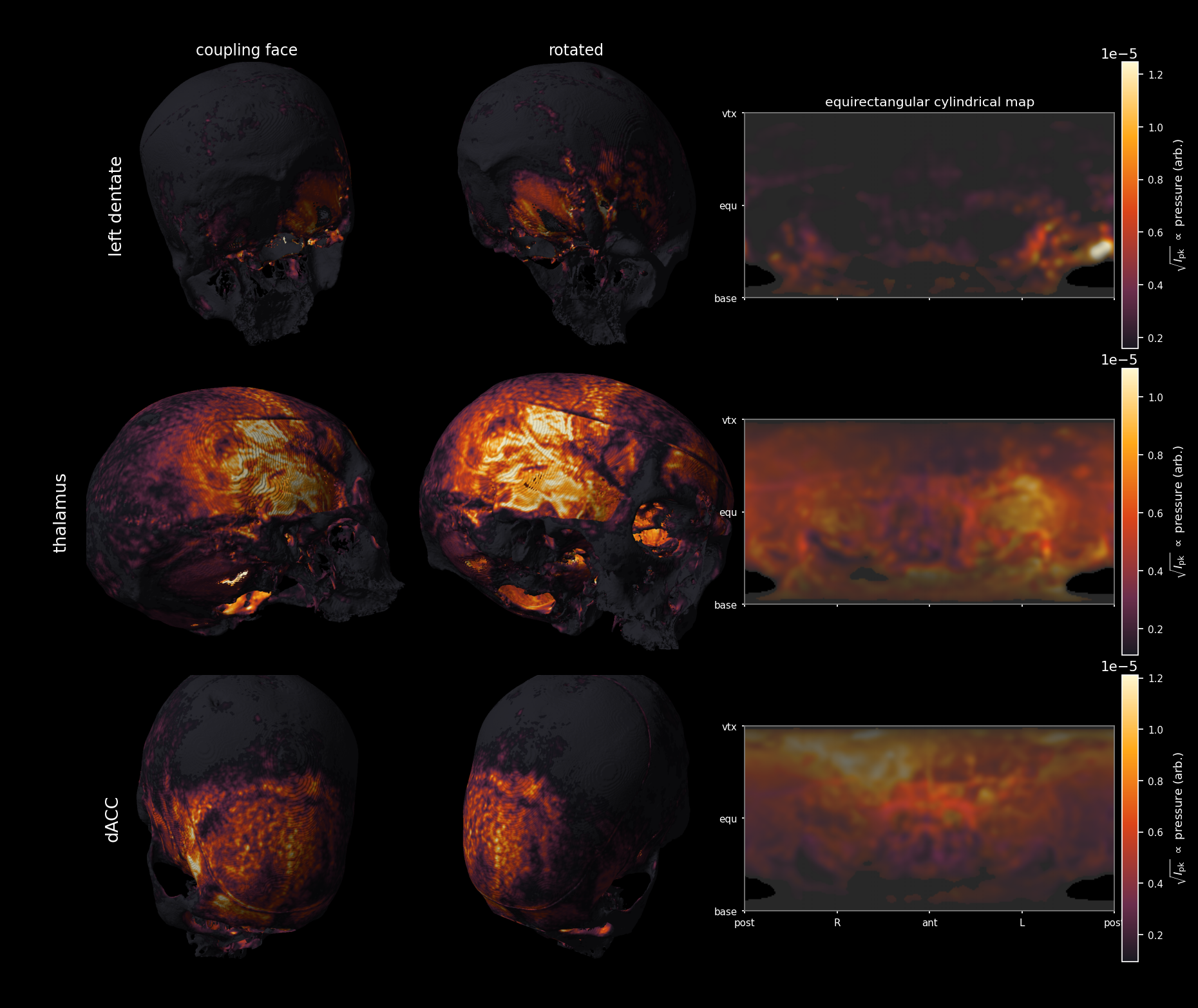}
\caption{Skull-surface coupling shown as $\sqrt{\text{peak intensity}}\propto$ pressure on the external
bone surface for the three targets (rows show the dentate, thalamus, and dACC), from the single \SI{6.16}{}-ppw
direct-arrival-windowed outward solve per target. The columns show two target-specific 3-D views
(coupling face, rotated) and a smooth equirectangular cylindrical map (superior pole, anterior-centred).
Each row on its own scale. The square root compresses the squared-intensity range so the distributed
coupling, not just the nearest-bone peak, is visible. For the dentate the foramen-magnum leak (a
near-zero-bone direct path, not a placeable window) is excluded by a no-bone ray test. The thalamus and
dACC, which couple through genuine distributed calvarium, are shown unmasked. The glow marks the
high-coupling window and the target-specific emergence (occiput for the dentate, broader calvarial coupling for the thalamus and dACC).}
\label{fig:transparency}
\end{figure}

To show the full pattern the coupling is rendered as $\sqrt{I_i}\propto$ pressure rather than
intensity (Fig.~\ref{fig:transparency}, each target on its own scale, with the square root compressing the squared-intensity range so the dim distributed coupling is visible). For the eccentric
dentate the brightest emergence is a direct path through the foramen magnum at the skull
base, a near-zero-bone leak rather than a placeable transcranial window. Excluding it with a
no-bone ray test (the same criterion applied to the focus in \S\ref{sec:targetfoci}) leaves the
genuine occipital coupling footprint. The central thalamus and anterior dACC have
no such hole leak, their low-bone coupling distributed over the calvarium (genuine windows,
not a single base aperture), so no exclusion is applied.

\subsection{Surface phase-aberration (delay) map}
\label{sec:delaymap}
\begin{figure}[t]
\centering
\includegraphics[width=\linewidth]{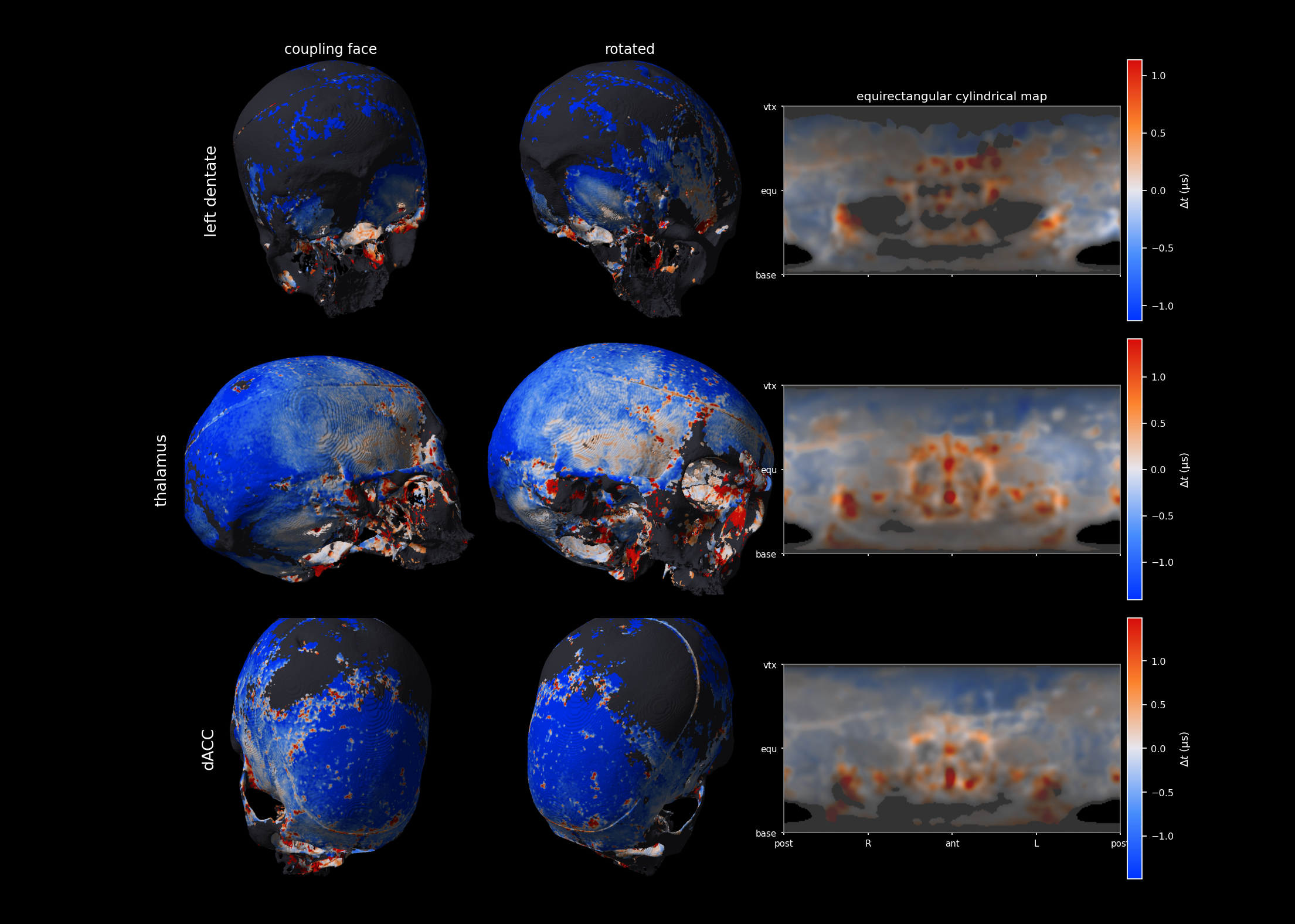}
\caption{Skull-surface arrival-time delay relative to spherical propagation, the unwrapped phase
aberration of Eq.~\eqref{eq:delay}, for each target, from its single \SI{6.16}{}-ppw outward solve.
The rows are the dentate, thalamus, and dACC. The columns show two target-specific 3-D views (coupling face,
rotated) and a smooth equirectangular cylindrical map (superior pole, anterior-centred). Blue = advanced
(fast through-bone path), red = retarded, dark grey = no coherent outward arrival (shadowed bone),
zero-median centred. Each row carries its own quantitative $\Delta t$ colorbar in \si{\micro\second}.}
\label{fig:delaysurf}
\end{figure}

The same outward solve gives the phase as a clean arrival-time map. Differencing the envelope-peak arrival time $t_{\mathrm{arr}}(\mathbf{x})$ (\S\ref{sec:cohmethod}) against the spherical
(homogeneous-medium) reference time gives
\begin{equation}
\Delta t(\mathbf{x}) \;=\; t_{\mathrm{arr}}(\mathbf{x}) \;-\; \frac{|\mathbf{x}-\mathbf{x}_t|}{c_0},
\label{eq:delay}
\end{equation}
zero-median centred over the coherent surface, with $\mathbf{x}_t$ the target (\S\ref{sec:validation})
and $c_0$ the reference sound speed. This $\Delta t$ is the per-element transmit-delay correction
expressed directly as a time, recovered for the entire skull surface from the same single solve
(Fig.~\ref{fig:delaysurf}). Expressed this way the aberration is unwrapped, carrying
no $2\pi$ ambiguity, unlike a single-frequency phase map, so it can be read and applied without phase
unwrapping. The delay advances (negative) where the path crosses more fast cortical bone and
retards where it grazes thicker or more oblique bone. Across the coherent coupling region the
zero-median-centred delay spans several microseconds ($\Delta t\approx-1.6$ to
$+\SI{3.3}{\micro\second}$ for the dentate, and comparably, $\approx-2$ to $+\SI{5}{\micro\second}$, for the thalamus and dACC), the
per-element transmit-delay correction the array applies. The map is shown only where the outward wave
arrives coherently (the accessible coupling region). Elsewhere the bone is shadowed.

\subsection{Aberration coherence length}
\label{sec:coherence}
\begin{figure}[t]
\centering
\includegraphics[width=\linewidth]{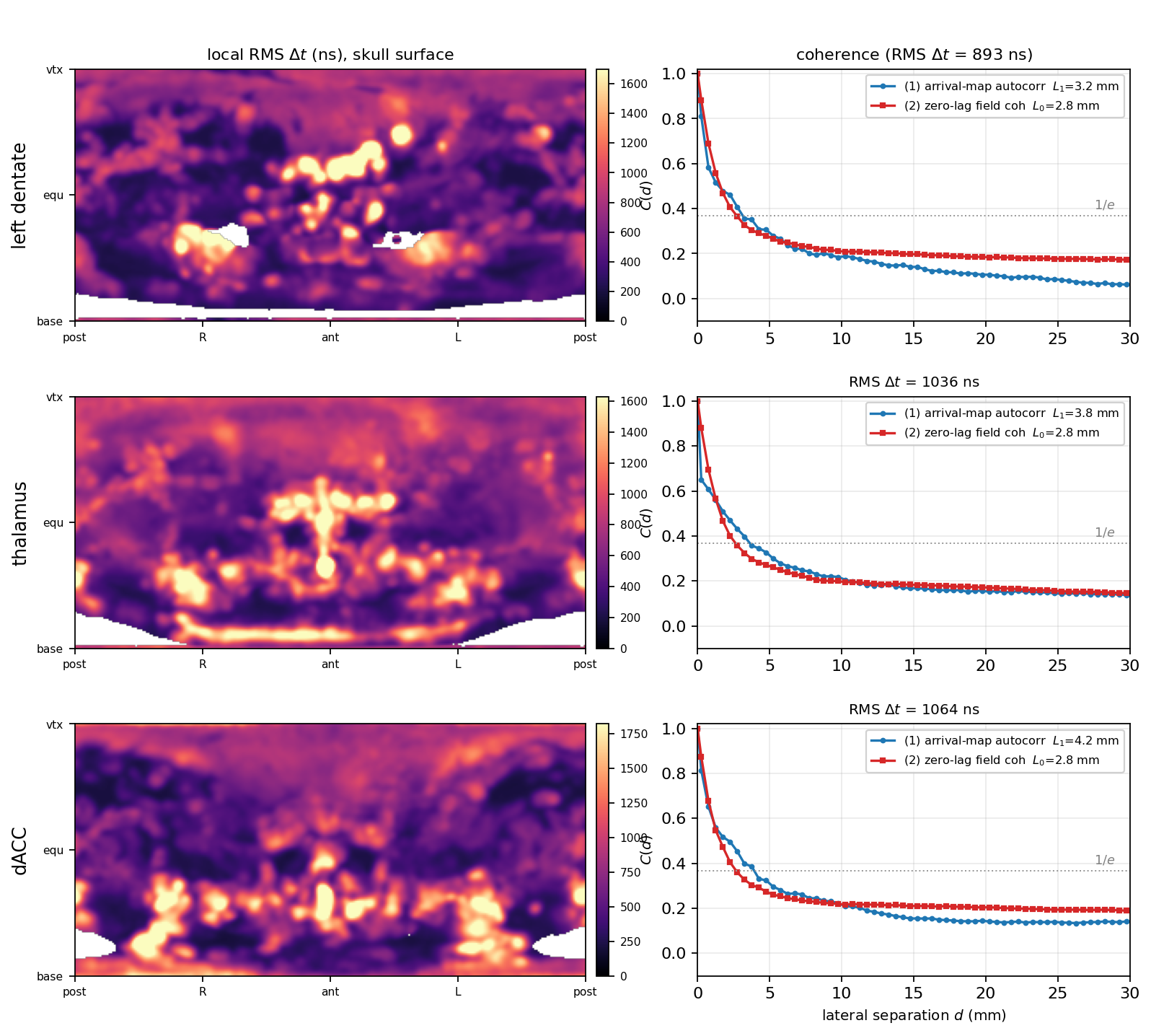}
\caption{Skull-surface arrival-time aberration per target (rows show the dentate, thalamus, and dACC), on the
\SI{6.16}{}-ppw graded skull at \SI{1}{\mega\hertz}, from the windowed direct-arrival field of the single
outward solve. The left column shows the local RMS delta-delay on the superior-pole equirectangular surface
(amplitude-weighted, neighbourhood RMS), largest over thick or complex bone and smallest over the thin,
uniform windows. The right column shows two coherence-length estimators versus lateral separation $d$, the
arrival-time-map autocorrelation $C_1$ ($L_1=3.25$ to \SI{4.25}{\milli\meter}) and the zero-lag field
(van~Cittert--Zernike) coherence $\rho_0$ ($L_0=\SI{2.75}{\milli\meter}$). The two estimators,
built from the scalar delay map and from the full waveforms respectively, corroborate each other. Both
are pinned to unity at $d=0$ by construction and cross $1/e$ (dotted) at the few-millimetre
phase-coherence scale. The RMS delta-delay ($0.89$ to \SI{1.06}{\micro\second}) is annotated per
panel.}
\label{fig:coherence}
\end{figure}

The single outward solve also measures the spatial coherence of the aberration, the property that helps
set the element pitch and bounds single-element focusing. The two estimators of \S\ref{sec:cohmethod} ($C_1$, $\rho_0$) are shown in Fig.~\ref{fig:coherence} and agree across the three targets. The delay-map autocorrelation
gives $L_1=3.25$, $3.75$, and \SI{4.25}{\milli\meter} for the dentate, thalamus, and dACC, the zero-lag
waveform coherence gives $L_0=\SI{2.75}{\milli\meter}$ for all three, and the RMS delay is $0.89$,
$1.04$, and \SI{1.06}{\micro\second} (about one period at \SI{1}{\mega\hertz}). Mapped over the surface
(Fig.~\ref{fig:coherence}, left), this RMS delay is largest over thick or complex bone and smallest over
the thin, uniform windows. The aberration phase thus stays
coherent over only a few millimetres. Because a broadband oscillatory pulse is inverted by a quarter-period
delay error, an untimed (zero-lag) sum loses coherence once the delay drifts by that much. This sets
two array requirements. The element pitch must resolve the few-millimetre coherence scale (the
$\le\!\lambda/2$ pitch is finer), and each element must be timed to the measured $\Delta t$ to correct
the $\sim\!\SI{1}{\micro\second}$ RMS delay, which a single rigid element spanning many coherence
lengths cannot (\S\ref{sec:arrayvssingle}). The same few-millimetre scale also sets the
placement tolerance. Displacing the aperture or mis-localizing the target beyond the
coherence length moves each element onto skull whose aberration no longer matches its recorded
correction, so positioning and registration need few-millimetre accuracy, tightest for targets reached
through thick or complex bone and loosest through thin, uniform windows.

\subsection{Transparency-guided transducer placement}
\label{sec:placement}
Searching the delivered-energy-optimal window over the whole skull (raw coupling,
incidence-weighted) places a focused transducer for the dentate on the suboccipital bone at MNI
$\approx[-15,-70,-57]$~mm, only \SI{27}{\milli\meter} from the target (\ang{29.4} incidence, 15~mm
footprint), a short, thin-bone posterior-inferior path through solid bone (the dentate-to-window
ray crosses cortical bone, with the foramen-magnum leak excluded, \S\ref{sec:transparency}). This
window is closer and more transparent than the best the occipital-array model could find
($[-11,-78,-51]$~mm on the occipital vault, \SI{44}{\milli\meter} away). Truncating the domain to
the instrumented vault hid the suboccipital approach that the whole-skull search recovers. This \SI{27}{\milli\meter}
suboccipital point is where the transparency map peaks. The pose-optimized 64 mm spherical transducer
(\S\ref{sec:surfint}) is then seated nearby at $[-31,-73,-55]$, \SI{33}{\milli\meter} out
(Table~\ref{tab:targets}), trading a little window-centre coupling for a foramen-clean full aperture.
The chosen placement is
exported in the planning pipeline's subject (NRRD-voxel-mm) frame with the focal length applied
after the (non-isometric) atlas-to-subject warp, so the focus lands exactly on the target.

\subsection{Generalization to other deep targets}
The procedure is target-agnostic (only the source location changes, while the skull, Fullwave~2, and registration are shared). Repeating it for two further deep targets returns distinct, anatomically-appropriate windows
(Table~\ref{tab:targets}) whose differences are themselves informative. The central thalamus is reached at near-normal
incidence through a lateral window, and incidence-legal skull subtends most of $4\pi$ around it (Table~\ref{tab:targets}).
A deep central target is reachable from almost anywhere, so
its placement is forgiving. The dentate, equally deep but eccentric in the posterior fossa, has
a comparably broad accessible area yet no window faces it normally, so its
optimum is obligately oblique, the constraint being incidence, not area. The dorsal
anterior cingulate (dACC), anterior and shallower, is reached through a broad lateral
region at near-normal incidence. Its small aperture is set not by
window area but by its short focal depth (\S\ref{sec:aperturesize}). The
single-element aberration penalty (\S\ref{sec:crossover}) grows with incidence, so it is largest for the eccentric
dentate and smallest for the near-normal thalamus, whereas the phased array, correcting
the aberration, is comparatively target-insensitive.

\begin{table}[hbt]
\centering
\caption{Transducer-objective targeting windows (per-element no-bone foramen drop, pose-optimized over the
64 mm aperture) for three deep targets, each from one outward solve. These are the seats that deliver
the verified foci of \S\ref{sec:targetfoci}. They refine the bare transparency ($\sqrt{J}$) optimum
(Fig.~\ref{fig:placementobj}), within a few mm for the dentate, \SI{18}{\milli\meter} inferior for the
thalamus, and \SI{10}{\milli\meter} superior for the dACC.
MNI mm. Incidence = true surface normal versus beam at the window. Access = the solid angle, as a
fraction of $4\pi$ about the target, subtended by good-transmission, incidence-legal (incidence at most
\ang{30}) skull, computed by binning the target-to-patch directions of the legal patches over the
sphere and summing the occupied-bin solid angle, with any foramen-magnum / no-bone leak excluded by the
ray test of \S\ref{sec:transparency} (its near-zero-bone paths are oblique and fall outside the
incidence-legal set, so the access fractions are unchanged by the exclusion).}
\label{tab:targets}
\begin{tabular}{lccccc}
\toprule
Target & Target (MNI) & Window (MNI) & Dist.\ (mm) & Incid. & Access \\
\midrule
Thalamus (L)  & $[-12,-18,8]$   & $[-67,-7,4]$    & 56 & \ang{3.2}  & $94\%$ \\
Dentate (L)   & $[-12,-57,-34]$ & $[-31,-73,-55]$ & 33 & \ang{29.4} & $77\%$ \\
dACC          & $[-4,24,28]$    & $[-57,41,24]$   & 56 & \ang{18.8} & $82\%$ \\
\bottomrule
\end{tabular}
\end{table}

\begin{figure}[tb]
\centering
\includegraphics[width=\linewidth]{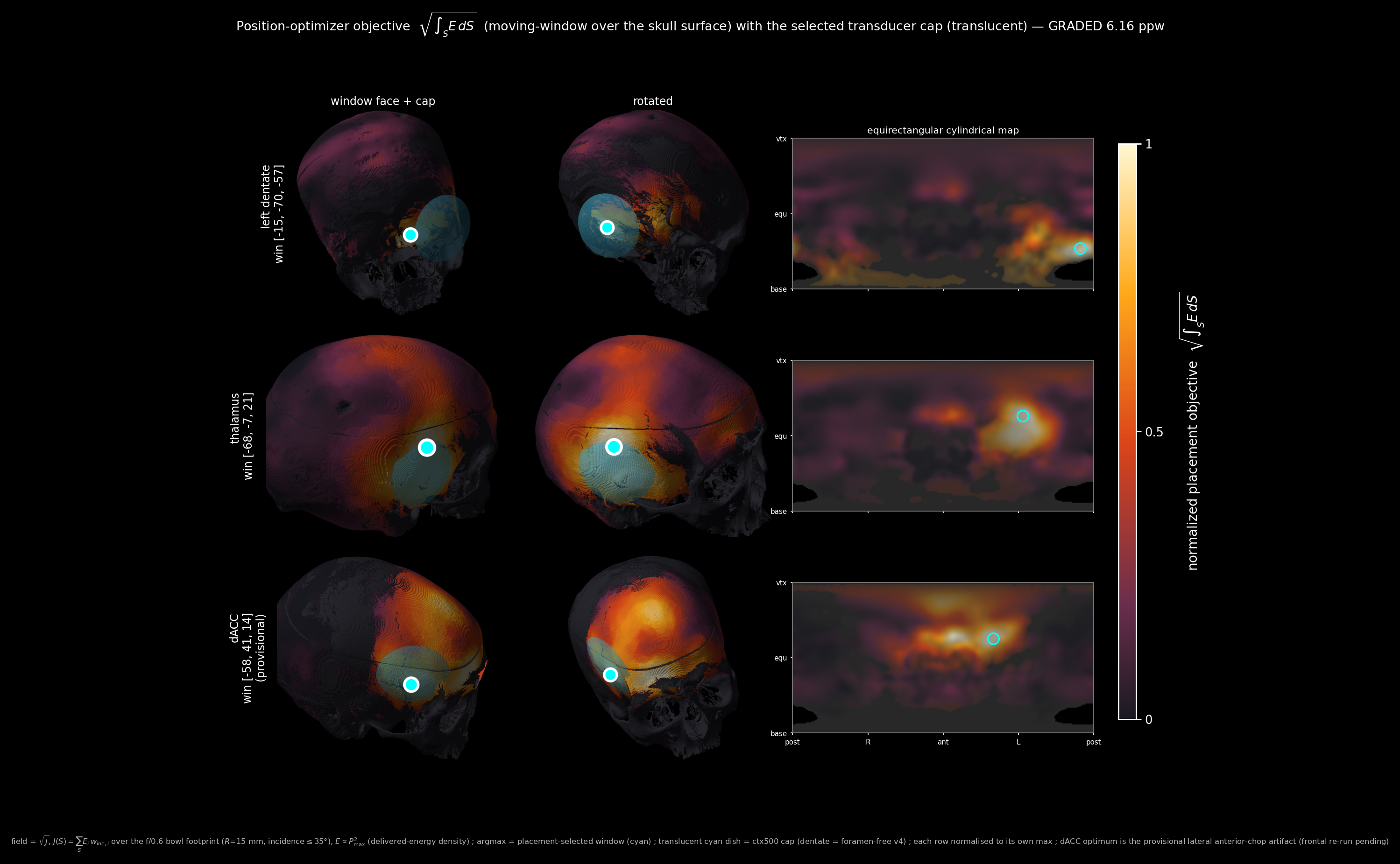}
\caption{The placement objective field on the skull, for the three deep targets (rows show the dentate,
thalamus, and dACC). The field shown is the array optimizer, the delivered-energy integral
$\sqrt{J(S)}$, one of the two optimizers read from the single solve (\S\ref{sec:arrayvssingle}). The
outer skull is coloured by the placement score $\sqrt{J(S)}$ of
\S\ref{sec:surfint}, the achievable focal peak, with $J(S)=\int_S E\,dS$ the delivered-energy
density $E$ integrated over the validated $f/0.6$ transducer footprint $S$ at each candidate window centre
(oblique patches incidence-down-weighted, as in the access criterion of Table~\ref{tab:targets}). It is shown
in two target-specific 3-D views (window face, rotated) and a smooth equirectangular cylindrical map,
matching Fig.~\ref{fig:transparency}. The translucent 64 mm spherical transducer is overlaid at its pose-optimized
targeting seat (Table~\ref{tab:targets}, \S\ref{sec:targetfoci}). The marker is the bare $\sqrt{J}$
field maximum (the coupling optimum), which the transducer-objective seat refines and can therefore sit off
(the thalamus seat is \SI{18}{\milli\meter} inferior, the dACC \SI{10}{\milli\meter} superior).
The bright lobe is broad and near-centred for the thalamus (forgiving, near-normal placement), broad
but obliquely displaced for the dentate (incidence-limited), and a more confined lateral region for
the dACC. The dACC panel is provisional, shown at its lateral window because its frontal approach
cannot be assessed with the present medium, which models intracranial air (notably the frontal sinus)
as water, so a frontal window would appear spuriously open (see Discussion). Colour is normalised per target to its
own maximum.}
\label{fig:placementobj}
\end{figure}

The placement objective behind these windows can be seen directly. Figure~\ref{fig:placementobj}
renders the placement score $\sqrt{J(S)}$ of \S\ref{sec:surfint} (Eq.~\ref{eq:surfint}) on the outer skull for each target, with the selected transducer overlaid
translucent. Unlike the raw per-patch coupling map (Fig.~\ref{fig:transparency}), this is the
optimizer's score at each candidate window position. Its bright maximum coincides, for all
three targets, with the analytic optimum of Table~\ref{tab:targets} (argmax-to-optimum
$\le\!0.4$~mm), so the single recorded field selects the placement with no per-candidate search. The
breadth of the lobe also reads off the placement margin of Table~\ref{tab:aperture}, broadest for the
forgiving near-normal thalamus, tightest for the dACC.

The same solve also bounds the optimal aperture size (\S\ref{sec:aperturesize}). All three targets' access (Table~\ref{tab:targets}) far exceeds the $\sim\!25\%$ solid-angle demand of an $f/0.6$ transducer ($3.1$--$3.8\times$ margin, Table~\ref{tab:aperture}), so each is practically (grating, fabrication) rather than window limited. The deep dentate and thalamus ($R\!\sim\!\SI{70}{\milli\meter}$) take the same large $D\!\approx\!\SI{12}{\centi\meter}$ transducer, while the shallow dACC ($R\!\sim\!\SI{25}{\milli\meter}$) shrinks to $D\!\sim\!\SI{4}{\centi\meter}$ at the same $f/0.6$, placing its focus in the near field (largest offset \SI{5.2}{\milli\meter}, \S\ref{sec:targetfoci}).

\begin{table}[hbt]
\centering
\caption{Optimal aperture size per target. The focusing $f$-number is fixed ($\sim\!f/0.6$, the validated \ang{120}
array, which subtends $\approx\!25\%$ of $4\pi$ at the focus), so $D\!\approx\!f_\#R$ scales with focal depth $R$. Margin $=$ access $\div$ the $\sim\!25\%$ demand. Access is the foramen-excluded solid angle of Table~\ref{tab:targets}. Deep-target $D$ is the simulated occipital array. The dACC $D$ is a geometric estimate.}
\label{tab:aperture}
\small
\begin{tabular}{lccc}
\toprule
Target & Access ($\Omega_\mathrm{good}/4\pi$) & Optimal $D$ (at $f/0.6$) & Margin \\
\midrule
Thalamus & $94\%$ & $\approx\!\SI{12}{\centi\meter}$ (large)            & $3.8\times$ \\
Dentate  & $77\%$ & $\approx\!\SI{12}{\centi\meter}$ (large, oblique)   & $3.1\times$ \\
dACC     & $82\%$ & $\sim\!\SI{4}{\centi\meter}$ (small, near-field)    & $3.3\times$ \\
\bottomrule
\end{tabular}
\end{table}

\subsection{Each selected window delivers an on-target focus}
\label{sec:targetfoci}
\begin{figure}[t]
\centering
\includegraphics[width=\linewidth]{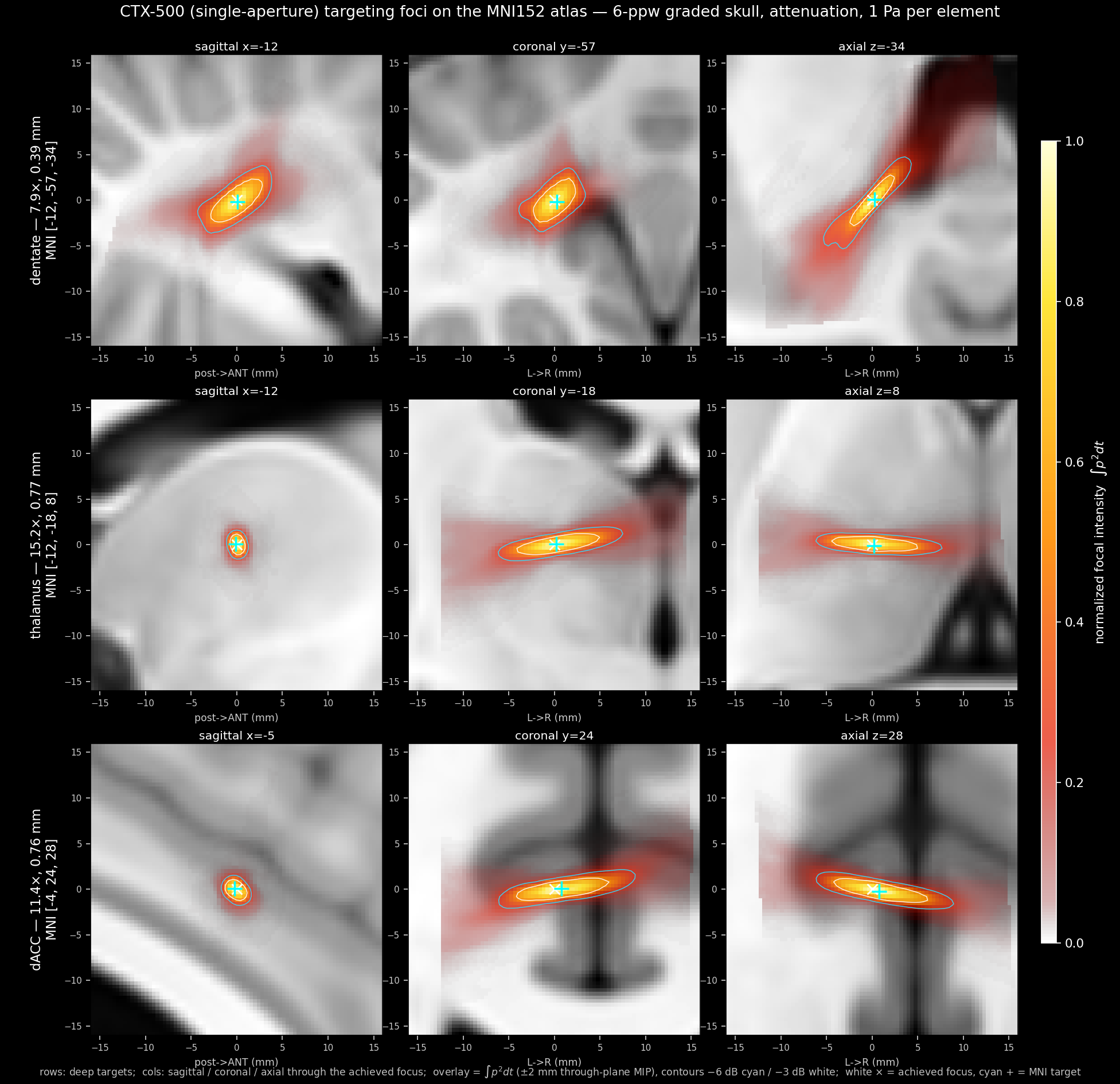}
\caption{Measured (time-reversal) focus at each target's optimal window, on the
MNI152 ICBM 2009a atlas. The rows are the left dentate (suboccipital window), thalamus
(lateral), and dorsal anterior cingulate (lateral). The columns are the sagittal, coronal, and axial cuts through the achieved focus.
For each target one outward solve selects the window. A rigid 64 mm spherical transducer (\SI{63.2}{\milli\meter} ROC, \SI{64}{\milli\meter} aperture) at that
window is driven by the time-reversed recorded field (each element \SI{1}{\pascal} peak) through the full skull, and the resulting focal
intensity ($\int p^2\,dt$) is overlaid ($\pm2$~mm through-plane maximum-intensity projection,
$-6$~dB cyan / $-3$~dB white contours, white $\times$ = achieved focus, cyan $+$ = MNI target). All three
land on target (thalamus $15.2\times$/\SI{0.77}{\milli\meter}, dACC $11.4\times$/\SI{0.76}{\milli\meter},
dentate $7.9\times$/\SI{0.39}{\milli\meter}). The dentate's transducer sits in the suboccipital basin, the
foramen-magnum leak removed by a per-element no-bone drop ($\sim$0.2\% of the aperture) rather than by
relocating the transducer. This is
distinct from the occipital-array focus of Fig.~\ref{fig:focus}, which validates the focusing model at
its instrumented vault window. Here each target is focused through its own build-able, foramen-clean window.}
\label{fig:targeting}
\end{figure}

Each transparency-selected window must actually deliver an on-target focus. For every target a rigid 64 mm spherical transducer was placed (\SI{63.2}{\milli\meter} radius of curvature, \SI{64}{\milli\meter} aperture, the geometry of a commercial CTX-500 but driven here at \SI{1}{\mega\hertz} for higher focal gain, with a matched \SI{500}{\kilo\hertz} study left to future work) at its window and re-emitted the time-reversed recorded field (each element peaking at \SI{1}{\pascal}) through the full skull, mapping the focal volume onto the MNI152 atlas (Fig.~\ref{fig:targeting}). All three foci land on target, the thalamus \SI{0.77}{\milli\meter} from its MNI target at $15.2\times$ ($-6$~dB FWHM $7.25\times1.75\times\SI{2.0}{\milli\meter}$), the dorsal anterior cingulate within \SI{0.76}{\milli\meter} at $11.4\times$ ($6.0\times2.25\times\SI{2.0}{\milli\meter}$), and the dentate within \SI{0.39}{\milli\meter} at $7.9\times$ ($2.75\times1.75\times\SI{4.5}{\milli\meter}$).

The dentate is the one target where the foramen magnum complicates the aperture, as a per-element rim effect rather than a window-selection constraint. A transducer placed by eye directly behind the nucleus leaks badly, with $\sim$24\% of its elements reaching the target through the foramen along zero-bone paths, but that opening is a low-impedance escape, not a placeable window. Aimed instead from the suboccipital window that the map already rates highest (the \SI{27}{\milli\meter} optimum), a full transducer clears the foramen almost on its own, dropping only $\sim$0.2\% of its elements ($\sim$80 of \num{49000}) by the no-bone ray test of \S\ref{sec:transparency}. It therefore stays in this basin, $\sim$\SI{33}{\milli\meter} out at \ang{29} incidence, and refocuses at $7.9\times$, nearly twice the gain at under half the distance of the \SI{79}{\milli\meter} occipital-squama seat ($4.05\times$) that a whole-transducer foramen-free constraint would force.

Pose-optimizing for this foramen-excluded delivered energy (\S\ref{sec:surfint}) improves all three targets. It seats the thalamus transducer \SI{18}{\milli\meter} inferior onto near-normal bone (\ang{3} versus \ang{14}), lifting the verified focus to $15.2\times$ from $11.8\times$, and shifts the dACC seat \SI{10}{\milli\meter} superior so the whole aperture crosses bone (firing every element rather than dropping the $\sim$8.5\% that would escape through the orbit), lifting it to $11.4\times$ from $10.4\times$. The same single outward solve that selects each window thus also predicts the achieved focus, closing the placement loop without a per-candidate search.

\subsection{A full-wave test of array versus single-element placement}
\label{sec:crossover}
The array-versus-single distinction (\S\ref{sec:arrayvssingle}) was tested on the whole skull with a full-wave ground truth,
a surface-conformal posterior-fossa receiving array on the whole-skull
model, and energy-matched inward re-simulations at two contrasting equal-size windows for the dentate
(a closer suboccipital window and the occiput, \SI{40}{\milli\meter} apart) under both the
array (time-reversal) and single-element (homogeneous-delay) drives. The dominant effect is
the aberration penalty, the single element delivering a $2.8$--$3.5\times$ lower focal peak than
the phased array at the same window, the aberration the array corrects and the single element cannot. The placement
sensitivities then differ as predicted, though weakly. Having corrected the aberration, the array is
nearly insensitive to which of the two windows it occupies ($1.04\times$, within noise). The single
element is window-sensitive and, lacking phase correction, favours the closer suboccipital window by
$1.22\times$. The two optima therefore differ, a weak crossover in the predicted direction (the array
optimizer to the occiput, the single-element optimizer to the suboccipital window),
but the effect is secondary to the aberration penalty, and on this posterior-fossa anatomy the occiput
is nearly optimal for both. As for the focusing gain, a single-frequency surrogate mis-ranks the broadband single-element preference, and the broadband re-simulation is the reference.

\subsection{Validation on the occipital-array model}
\label{sec:surfintresult}
The multi-element results use the occipital-array focusing model (\S\ref{sec:skullmodel}), the
only model with recorded per-element signals. Its instrumented window is the occipital vault
($\approx[-12,-76,-52]$~mm), distinct from the whole-skull suboccipital transducer window of
\S\ref{sec:placement}. To test the placement objective at a fixed element budget, the framework drives $N=64$ recorded element
sites and compares their placements, at matched $N$ and matched
transmit power (\S\ref{sec:validation}).
Because the on-target peak under time reversal scales as $\sqrt{\sum_i E_i}$, placement value
is the total delivered energy. Against an already-good, geometry-chosen occipital window
populated uniformly, the map's additional refinement is $1.4\times$ energy
($1.2\times$ peak). Once the correct window is found, time reversal co-phases the elements
regardless of their exact placement, so the map's value is largest in choosing the
window and for targets whose geometrically-obvious window couples poorly. Separately, at a
fixed placement, focusing with the medium's true (aberration-corrected) delays raises the
on-target peak $2.7\times$ over homogeneous-medium geometric delays. This is the skull aberration that time reversal corrects, and it dominates the
multi-element gain.

The aberration result was confirmed with a full-wave inward re-simulation of the rigid \ang{120}
occipital array under both drive laws, energy-matched, recording a full-resolution
(\SI{0.375}{\milli\meter}) focal box at the dentate. Time reversal produces a
$7.5\times$ higher peak pressure ($56\times$ peak intensity) than geometric delay focusing, and the
two foci differ qualitatively. Time reversal lands within \SI{0.4}{\milli\meter} of the target with a
near-wavelength spot ($1.1\times1.1\times\SI{0.8}{\milli\meter}$ FWHM), whereas geometric focusing
both mis-steers the focus by \SI{4.3}{\milli\meter} (the skull bends the beam) and broadens it. The
array's per-element phase correction is what recovers a tight, on-target focus through the occipital
bone. Geometric (single-element-equivalent) focusing scatters most of that energy away from the
target.

The surface-integral predictor was then tested against this same ground truth. Evaluated over candidate windows, the objective
$\sqrt{J(S)}$ (Eq.~\ref{eq:surfint}) is maximal on the occipital/suboccipital calvarium
(MNI $\approx[-12,-76,-52]$~mm, 20~mm aperture), the anatomically expected approach, recovered by a single
moving-transducer pass over the precomputed map. Its predicted relative placement gains match the
recorded-signal ground truth to within $1\%$ (e.g.\ $1.19\times$ focal peak for the
transparency window versus a uniform tiling of the same window, reproducing the
recorded-signal placement ladder exactly, since the discrete $\sqrt{\sum_i E_i}$ is the
sampled $\sqrt{J}$). The angular-spectrum point-spread, computed from the same single solve with
no further simulation, predicts a transverse focal FWHM of
$\SI{2.9}{}\times\SI{3.0}{\milli\meter}$ against the inward re-simulation's
$\SI{2.9}{}\times\SI{2.8}{\milli\meter}$ and correctly returns a long
axial depth of field. The phase-conjugate drive is the verified power-constrained optimum (no
random unit-power drive exceeds it over $10^3$ trials), and the radial projection round-trips and
reproduces a free-space point source to machine precision.

Consistent with \S\ref{sec:surfint}, the stated caveats hold in the data, the single-frequency phase-conjugate-versus-geometric ratio being $\sim$\num{11}$\times$, far above
the broadband/measured \num{2.7}--\num{3.0}$\times$ focusing gain, and the single-frequency axial
depth of field overestimated ($\sim\SI{33}{}$ vs $\sim\SI{25}{\milli\meter}$). Both affect the
focusing predictions, not the energy-integral placement, which matched throughout.

\FloatBarrier
\section{Discussion}
The reciprocity argument underlying the whole-skull map holds for
linear, heterogeneous media, and at neuromodulation pressures this is,
generally, well satisfied. However a limitation lies in the medium
itself. The present skull model assigns water sound speed to every
non-bone voxel, so intracranial air (the paranasal sinuses, mastoid
air cells and airway) is not represented, and any window whose path
crosses such a cavity is modelled as spuriously transparent. This
bears most on anterior targets, where the frontal sinus lies directly
in the natural frontal approach, so the dACC is reported through a
lateral window, deferring a faithful assessment of frontal access
until intracranial air is segmented and acoustically modelled in the
medium pipeline.

A related limitation is the fluid wave model, which propagates only the longitudinal mode. Beyond the
water-to-bone longitudinal critical angle (near \ang{30} for the speeds used here, \cite{clement2004,white2006})
the longitudinal wave ceases to propagate and transmission proceeds by mode conversion to the bone
shear mode, which the solver does not represent. The \ang{30} incidence limit keeps the analysis within
the longitudinal regime, and because the omitted shear pathway can itself transmit at oblique
incidence \cite{clement2004}, the model is conservative at the window edge rather than over-optimistic.
All three selected windows, including the dentate at \ang{29.4}, lie within this limit.

The same machinery extends from single-transducer window selection to multi-element aperture design, where the per-site coupling ranks and apodizes the elements that densely populate the chosen window, with the inward simulation as the coherence check.
Relative to existing practice, the transparency map replaces a per-candidate search (in which each
transducer window is scored by its own full-wave transmit solve or by a single-frequency phase
surrogate) with a single outward solve whose surface integral is reused across all candidate
placements. This changes the cost of whole-skull placement from one wave simulation per candidate to
one simulation per target, with the remaining per-candidate work reduced to a moving-window surface
integral evaluated in seconds on the precomputed field. That one remaining per-target solve is not tied to
the finite-difference solver used here: a companion angular spectrum method \cite{pinton2026asm}
reproduces the Fullwave~2 focal-plane intensity to 1.1\% RMS through an ex vivo human skull and, for a
focused transducer, runs with $9\times$ less memory and $2.9\times$ less wall time, so the same surface
field could be produced at lower cost. Because each downstream use is a re-weighting
of the same surface field rather than a new wave solve, the precomputed map extends naturally to array
apodization for grating-lobe control, multi-target scheduling, and registration-robust placement.  The
selected aperture and its recorded per-element drives also feed directly into end-to-end models of
therapeutic effect, such as the companion open-source ultrasound-neuromodulation framework
\cite{pinton2026neuromod}, which consumes exactly this class of transcranial field solution.

\section{Conclusion}
This work has shown that a single full-wave time-reversal solve, the same solve already required for
aberration-corrected focusing, yields the transmit coupling of the entire skull surface for a fixed
target, a reciprocity-based skull transparency map. From this one map the delivered-energy-optimal
window, aperture and orientation follow by a moving-window surface integral, so transcranial
placement becomes inexpensive post-processing rather than a per-candidate wave solve. For the left
dentate nucleus the map selects a suboccipital window \SI{27}{\milli\meter} from the target at
\ang{29.4} incidence, where a full transducer clears the foramen magnum on its own and refocuses
on-target (\SI{0.39}{\milli\meter}) at $7.9\times$, while the same procedure returns readily-accessible
lateral windows that focus the thalamus and dorsal anterior cingulate at $15.2\times$ and $11.4\times$. On recorded per-element signals at matched element
count and transmit power, the single-solve surface-integral objective reproduces the per-candidate focal peaks to within tolerance.

It should be noted that the score ranks single-element coupling, which under time reversal is
exactly the quantity that governs the on-target peak, but it does not by itself constrain the focal
volume, off-target lobes, or the behaviour under non-ideal focusing. Those still call for the single
inward re-simulation that the pipeline performs. The sampled surface field is a relative coupling
proxy, so absolute pressure and the standing-wave loading of a physical, gel-coupled transducer
require a loaded solve. Within these bounds the method unifies the single-element coherent score of
prior reciprocity-based placement \cite{park2022} with a closed-form phased-array energy limit, and keeps
the full-wave physics of per-candidate planners at the cost of one solve. The same single-solve map should support array apodization for grating-lobe control,
multi-target scheduling, and registration-robust placement as natural extensions.

\appendix

\section{Interactive placement-positioning tool}
\label{sec:tool}

\begin{figure}[hbt]
\centering
\includegraphics[width=\linewidth]{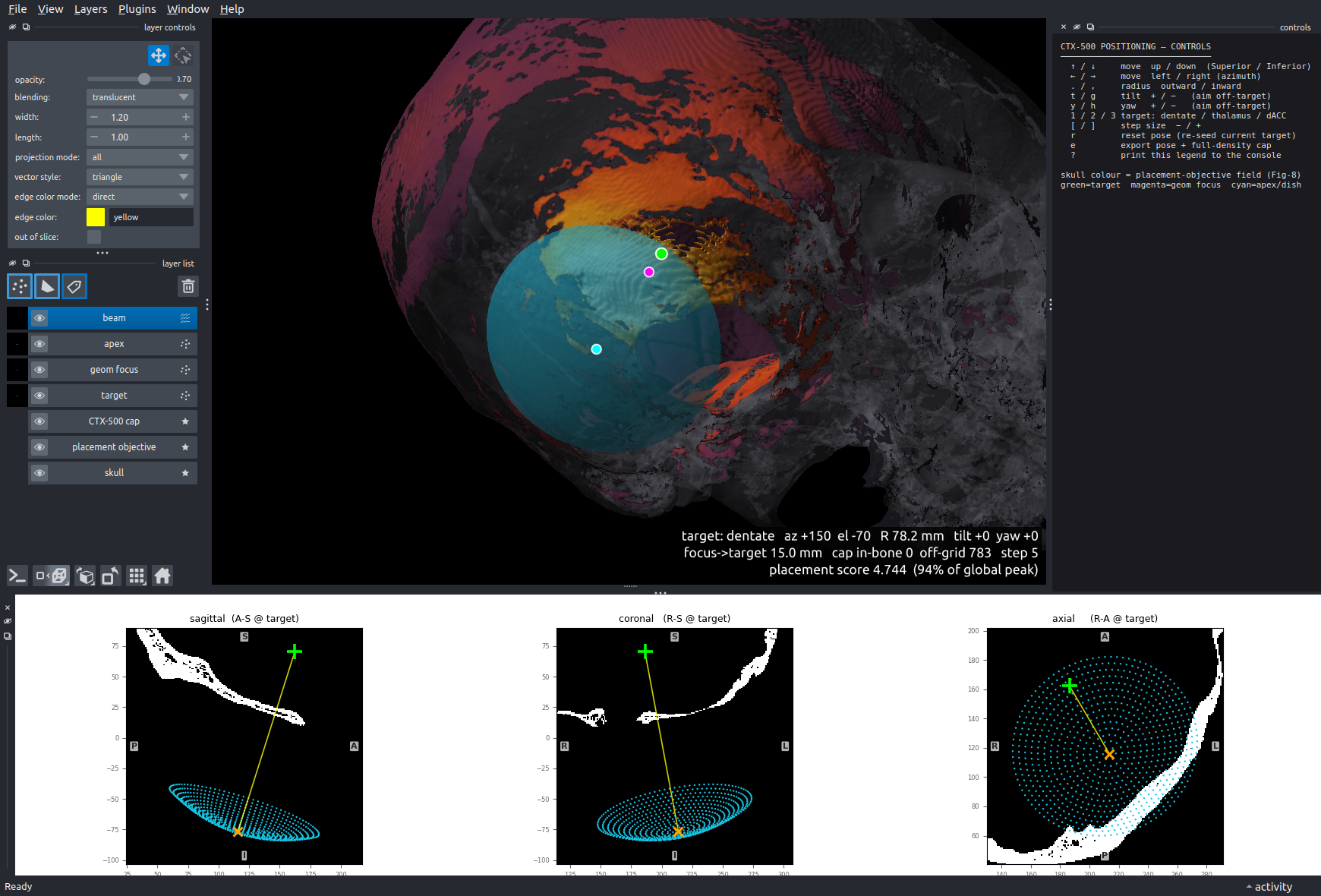}
\caption{Interactive placement-positioning tool, shown for the dentate. The centre panel shows the skull surface
coloured by the placement-objective field $\sqrt{J_\mathrm{w}}$ (\S\ref{sec:surfint}, dark to bright is
low to high single-element coupling), with a translucent 64 mm spherical transducer (cyan) seated on the occipital
window, the target (green) and its geometric focus (magenta), and the beam axis. The bottom panels show the same
placement in three orthogonal slices through the target (sagittal, coronal, axial), with the transducer
footprint (cyan), the target (green ``+'') and the transducer apex (orange) overlaid. The right panel lists the keyboard
controls. The on-canvas readout gives the live pose, the focus-to-target offset, the transducer clearance
from bone, and the placement score with its percentage of the global peak (here 94\% for a seated
suboccipital placement at \SI{15}{\milli\meter} standoff).}
\label{fig:tool}
\end{figure}

Because the placement objective of \S\ref{sec:surfint} is candidate-independent post-processing of a
single outward solve, the placement can be explored by hand in real time. This work provides a lightweight
tool (Fig.~\ref{fig:tool}) that lets an operator pose a rigid 64 mm spherical transducer (\SI{63.2}{\milli\meter}
radius of curvature, \SI{64}{\milli\meter} aperture) on any of the three targets and reads out the
same score the automated optimum maximises. The transducer face-centre rides on a sphere about the target.
The arrow keys slide it in azimuth and elevation, two keys change the standoff radius, and the transducer
points at the target by default, with optional tilt and yaw to steer the axis off-target. Targets are
switched with the number keys. Because the surface coupling map is precomputed, every pose is scored
without a new wave solve, so the readout updates instantly as the operator moves the transducer.

The skull surface is coloured by the placement-objective field (the moving-window $\sqrt{J_\mathrm{w}}$
of \S\ref{sec:surfint}, whose argmax is the selected window), with the translucent transducer, the target,
the geometric focus and the beam axis overlaid in three dimensions and the same scene shown in three
orthogonal slices through the target. A live panel reports the pose, the geometric-focus-to-target
offset, the number of transducer elements intersecting bone (a coupling-clearance check), and the placement
score together with its percentage of the global peak over the whole legal surface. The seated
suboccipital placement confirms the analytic optimum of Table~\ref{tab:targets} interactively (94\% of
the global peak for the dentate), and tilting or yawing the axis off-target lowers the score, making
the incidence acceptance constraint tangible. The posed transducer element coordinates and the full pose can
be exported to drive the inward time-reversal re-simulation of \S\ref{sec:placement}.

For accuracy the incidence-weight surface normals are obtained by local principal-component analysis of
the dense surface point cloud, which is free of the voxel-grid staircasing of a binarized-mask
gradient. The renderer uses napari for GPU three-dimensional display, and the whole tool consumes only
the precomputed surface map and skull mesh, not the multi-gigabyte medium.

\funding{This work was supported in part by NIH grants R01-EB037345 and
R01-EB036295, and by the Raynor Cerebellum Project.}

\data{The implementation of the whole-skull transparency map, the
surface-integral placement optimizer, and the interactive placement tool is
openly available at
\url{https://github.com/pinton-lab/skull_transparency}.}

\bibliographystyle{IEEEtran}
\bibliography{references}

\end{document}